# Revisiting the measurements and interpretations of DLVO forces

Bo Feng[1], Xiantang Liu[2], Xinmin Liu[1], Yingli Li[1], Hang Li[1]*

[1]*Chongqing Key Laboratory of Soil Multi-Scale Interfacial Processes, College of Resources and Environment, Southwest University, Chongqing 400715, P.R. China*
[2]*State Key Laboratory of Pollution Control and Resource Reuse, School of the Environment, Nanjing University, Nanjing 210023, P.R. China*

*Corresponding author at: Chongqing Key Laboratory of Soil Multi-Scale Interfacial Processes, College of Resources and Environment, Southwest University, Chongqing 400715, PR China.
E-mail address: lihangswu@163.com, lihang88@swu.edu.cn (H. Li).
Phone: 086-13883320589, 086-023-68251504.

Bo Feng (https://orcid.org/0000-0002-9861-3256)
Xiantang Liu (https://orcid.org/0009-0001-6120-9137)
Hang Li (https://orcid.org/0000-0002-8486-6631)

**Abstract:**

The DLVO theory and electrical double layer (EDL) theory are the foundation of colloid and interface science. With the invention and development of surface forces apparatus (SFA) and atomic force microscope (AFM), the measurements and interpretations of DLVO forces (i.e., mainly measuring the EDL force (electrostatic force) $F_{EDL}$ and van der Waals force $F_{vdW}$, and interpreting the potential $\psi$, charge density $\sigma$, and Hamaker constant $H$) can be greatly facilitated by various surface force measurement techniques, and would have been very promising in advancing the DLVO theory, EDL theory, and colloid and interface science. However, although numerous studies have been conducted, pervasive anomalous results can be identified throughout the literature, main including: (1) the fitted $\psi/\sigma$ is normally extremely small ($\psi$ can be close to or (much) smaller than $\psi_\zeta$ (zeta potential)) and varies greatly; (2) the fitted $\psi/\sigma$ can exceed the allowable range of calculation; and (3) the measured $F_{vdW}$ and the fitted $H$ vary greatly. Based on rigorous and comprehensive arguments, we have reasonably explained the pervasive anomalous results in the literature and further speculated that, the pervasive anomalous results are existing but not noticed and questioned owing to the two important aspects: (1) the pervasive unreasonable understandings of EDL theory and (2) the commonly neglected systematic errors. Consequently, we believe that the related studies have been seriously hampered. We therefore call for re-examination and re-analysis of related experimental results and theoretical understandings by careful consideration of the EDL theory and systematic errors. On these bases, we can interpret the experimental results properly and promote the development of EDL theory, colloid and interface science, and many related fields.

# Abbreviations

| | |
|---|---|
| $\Delta D$ | distance offset ($= D_0 - D'$) |
| $\delta$ | distance of the origin plane of the diffuse layer from the charged surface |
| $\varepsilon$ | relative permittivity |
| $\varepsilon_0$ | dielectric permittivity of vacuum |
| $\theta$ | equivalent tip cone half angle |
| $\theta_0$ | effective (actual) $\theta$ |
| $\theta'$ | apparent $\theta$ |
| $\kappa^{-1}$ | Debye length |
| $\pi$ | ratio of the circumference of a circle to its diameter |
| $\sigma$ | charge density obtained by directly fitting the measured $F_{EDL}$–$D$ curves, also the apparent diffuse layer charge density, which is equivalent to extrapolate $\sigma_\delta$ to the surface ($x = 0$) |
| $\sigma_\delta$ | charge density at the origin plane of the diffuse layer, also the diffuse layer charge density |
| $\sigma_s$ | surface charge density |
| $\sigma_x$ | charge density at distance $x$ from a charged surface |
| $\sigma_0$ | actual $\sigma$ |
| $\sigma'$ | misinterpreted $\sigma$ |
| $\sigma_{RMS}$ | root-mean-square (RMS) roughness |
| $\sigma_{std}$ | standard deviation of Gaussian distribution |
| $\psi$ | potential obtained by directly fitting the measured $F_{EDL}$–$D$ curves, also the apparent diffuse layer potential, which is equivalent to extrapolate $\psi_\delta$ to the surface ($x = 0$) |
| $\psi_\delta$ | potential at the origin plane of the diffuse layer, also called the diffuse layer potential |
| $\psi_\zeta$ | zeta potential |
| $\psi_m$ | midplane potential |
| $\psi_s$ | surface potential |
| $\psi_x$ | potential at distance $x$ from a charged surface |
| $\psi_0$ | actual $\psi$ |
| $\psi'$ | misinterpreted $\psi$ |
| A | a self-defined constant |
| $c$ | molar concentration of the electrolyte solution |
| $D$ | distance between two charged surfaces |
| $D_0$ | effective (actual) $D$ |
| $D'$ | apparent $D$ |
| $d_\zeta$ | distance of the hypothetical slip plane or shear plane from the charged surface |
| $E_\theta$ | error coefficient of $\theta$ |
| $E_\theta - 1$ | relative error of $\theta$ |
| $E_k$ | error coefficient of $k$ |
| $E_k - 1$ | relative error of $k$ |
| $E_R$ | error coefficient of $R$ |
| $E_R - 1$ | relative error of $R$ |
| e | base of the natural logarithm |
| $F_{DLVO}$ | DLVO forces |
| $F_{EDL}$ | EDL (or electrostatic) force |
| $F_{vdW}$ | van der Waals (or dispersion) force |
| $H$ | Hamaker constant |
| $H_0$ | actual $H$ |
| $H'$ | misinterpreted $H$ |
| $H_{p-v}$ | peak-to-valley height |
| $k$ | spring constant (or stiffness) of the spring devices |
| $k_0$ | effective (actual) $k$ |
| $k'$ | apparent $k$ |
| $k_B$ | Boltzmann constant |
| ln | natural logarithm |
| log | logarithm to any base |
| $N_A$ | Avogadro constant |
| $n_{i0}$ | bulk number concentration of ion $i$ in the electrolyte solution |
| $q$ | elementary charge |
| $R$ | equivalent radius or equivalent tip end radius |
| $R_0$ | effective (actual) $R$ |
| $R'$ | apparent $R$ |
| $T$ | absolute temperature |
| $t$ | Eq (3) |
| $x$ | distance from a charged surface |

**Figure captions**

# 1 Introduction
## 1.1 Importance of measurements and interpretations of DLVO forces

The Derjaguin–Landau–Verwey–Overbeek (DLVO) theory and electrical double layer (EDL) theory are the foundation of colloid and interface science and many other fields. The DLVO theory has recognized that the net interaction force between two adjacent charged surfaces in an aqueous electrolyte solution, the net DLVO force ($F_{DLVO}$), is decomposed into contributions from the two DLVO forces, EDL (or electrostatic) force, $F_{EDL}$, and van der Waals (or dispersion) force, $F_{vdW}$ (Fig. 1b) [1–3]. On this basis, through the measurements and interpretations of DLVO forces (i.e., mainly measuring the $F_{EDL}$ and $F_{vdW}$, and interpreting the potential $\psi$, charge density $\sigma$, and Hamaker constant $H$), the quantitative relationship between the microscopic interface properties (like EDL properties, including $\psi/\sigma$) and the mesoscopic and macroscopic processes (like colloidal stability and self-assembly) can be established (Fig. 1a–c) [2,3]. While with the invention and development of surface forces apparatus (SFA) [4–8] and atomic force microscope (AFM) [9–11], direct measurements and interpretations of DLVO forces have been greatly facilitated by various surface force measurement techniques, mainly SFA, AFM microsphere (colloidal) probe technique, and AFM nanoscale tip technique [12,13] (Fig. 1d–g).

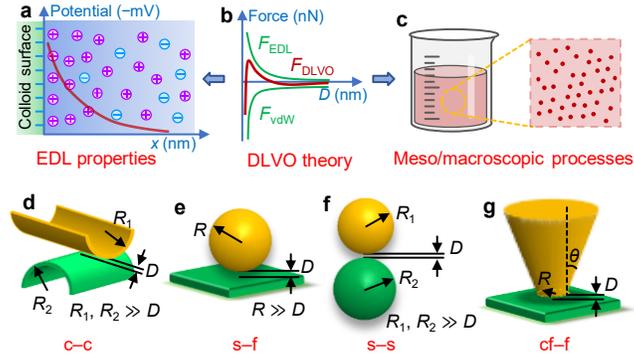

Fig. 1 Schematic of significance and techniques of measurements and interpretations of DLVO forces.

(**a–c**) Schematic of significance of measurements and interpretations of DLVO forces. (**a**) Microscopic properties of the EDL. (**b**) The DLVO forces obtained by surface force measurements, acting as a bridge between the microscopic and meso/macroscopic scales. (**c**) Meso/macroscopic processes.

(**d–g**) Interaction models used in different surface force measurement techniques. (**d**) The two-macroscopic-cylinder interaction model (denoted by c–c), with two macroscopic cylinders of radii $R_1$ and $R_2$ crossed at 90°, used in SFA. (**e**) The microsphere–flat surface interaction model (s–f), with a microsphere of radius $R$ and a flat surface, used in AFM microsphere probe technique. (**f**) The two-microsphere interaction model (s–s) with two microspheres of radii $R_1$ and $R_2$, used in AFM microsphere probe technique. (**g**) The conical tip with flat circular

end–flat surface interaction model (cf–f), with a tip of equivalent tip end radius $R$ and equivalent tip cone half angle $\theta$, and a flat surface, used in AFM nanoscale tip technique.

## 1.2 Pervasive anomalous results in the literature

However, although numerous studies have been conducted in measurements and interpretations of DLVO forces, pervasive anomalous results can be identified throughout the literature. The pervasive anomalous results mainly include: (1) the fitted $\psi/\sigma$ is normally extremely small ($\psi$ can be close to or (much) smaller than $\psi_\zeta$ (zeta potential)) and varies greatly; (2) the fitted $\psi/\sigma$ can exceed the allowable range of calculation; and (3) the measured $F_{vdW}$ and the fitted $H$ vary greatly. The details are described as follows.

(1) The fitted $\psi/\sigma$ is normally extremely small and varies greatly.

In existing literature, the fitted $\psi$ under different conditions ($c$ or pH) are commonly extremely small ($|\psi|$ is commonly < 100 mV, or even < 50 mV) for various materials, such as mica [14,15], silica [11,12,16–21], and polystyrene latex particles functionalized with sulfate [22–25], carboxyl [26] or amidine [24,25] groups. Moreover, the fitted $\psi$ can even be close to or much smaller than $\psi_\zeta$, such as silica [11,17,19,20], polystyrene latex particles [22–26], mica [27–29], etc.. Likewise, the extremely small fitted $\psi$ also reflects the extremely small fitted $\sigma$ (directly demonstrated in above literature or calculated by Eq. (7)), which is even only a few mC m$^{-2}$ (especially at lower $c$) and much smaller than surface charge density ($\sigma_s$). Besides, the fitted $\psi/\sigma$ normally varies greatly (under similar experimental conditions) as can be found in above literature. The typical examples for mica and silica are presented in Fig. 2. These phenomena have also been pointed out by other researchers [2,12,30]. In fact, as explained in section 2.6, the $\psi/\sigma$ can be relatively large (around $|\psi| \geq 150$ mV or $|\sigma| \geq 0.05$ C m$^{-2}$) at appropriate $c$ and pH.

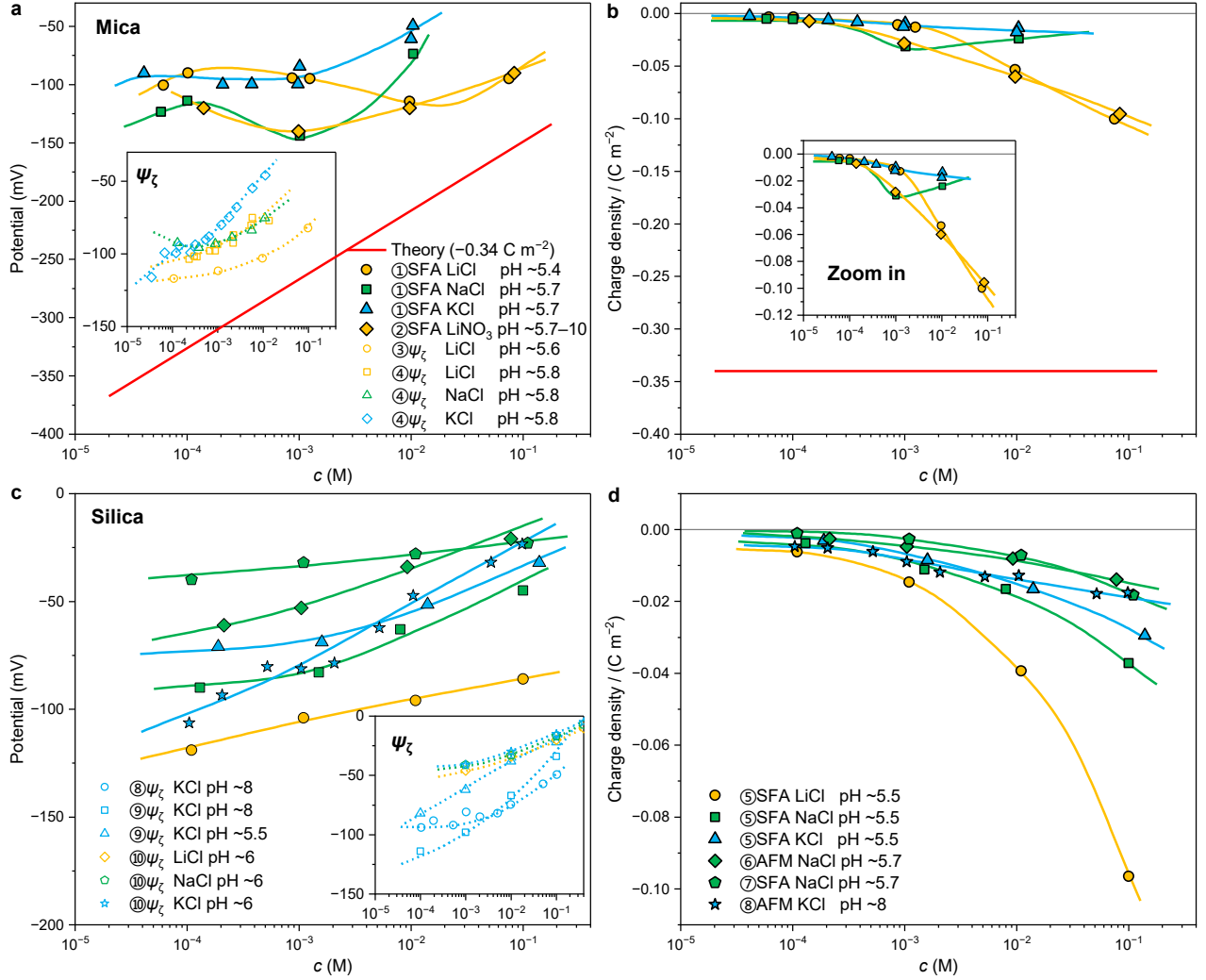

Fig. 2 The fitted $\psi/\sigma$ and the measured $\psi_\zeta$ for (**a**–**b**) mica and (**c**–**d**) silica. For mica, theoretical $\sigma$ is $-0.34$ C m$^{-2}$ and theoretical $\psi$ is calculated by Eq. (7).

The lines serve to guide the eye only. Data sources: 1 [14], 2 [15], 3 [29], 4 [27], 5 [18], 6 [17], 7 [16], 8 [20], and 9 [31], and 10 [32].

(2) The fitted $\psi/\sigma$ can exceed the allowable range of calculation.

As reviewed in our previous work [33], some over-approximated $F_{EDL}$ formulas (only valid at extremely small $\psi$ ($|\psi| < 25$ mV)) are frequently used in the literature. It is easy to find that, the misinterpreted (over-approximated) $\psi$ (denoted by $\psi'$) can seriously underestimate the actual $\psi$ (denoted by $\psi_0$), and $|\psi'|$ has a maximum value of $\frac{4k_BT}{q}$ (103 mV) as $|\psi_0|$ increases (Fig. 3a–b,

$\psi' = \frac{4k_BT}{q} \tanh\left(\frac{q\psi_0}{4k_BT}\right)$, refer to Eq. (S1) in ref. [33]). This means that it is obviously unreasonable for $|\psi'|$ to exceed 103 mV or even reach 150 mV, which however have been usually reported in the literature [34–36]. Similarly, the misinterpreted (over-approximated) $\sigma$ (denoted by $\sigma'$) can also seriously underestimate the actual $\sigma$ (denoted by $\sigma_0$) (especially at lower $c$) and $|\sigma'|$ has the maximum values of $\frac{4\varepsilon\varepsilon_0 k_B T\kappa}{q}$ (2.4, 7.5, 24 and 75 mC m$^{-2}$ at representative $c$ of $10^{-4}$, $10^{-3}$, $10^{-2}$ and $10^{-1}$ M) as $|\sigma_0|$ increases (Fig. 3c–d, $\sigma' = \frac{4\varepsilon\varepsilon_0 k_B T\kappa}{q}\tanh\left(\frac{1}{2}\mathrm{asinh}\left(\frac{q\sigma_0}{2\kappa\varepsilon\varepsilon_0 k_B T}\right)\right)$, refer to and Eq. (S2) in ref. [33]). The unreasonable cases of $|\sigma'|$ exceeding its maximum values can also be found in the literature [34,37].

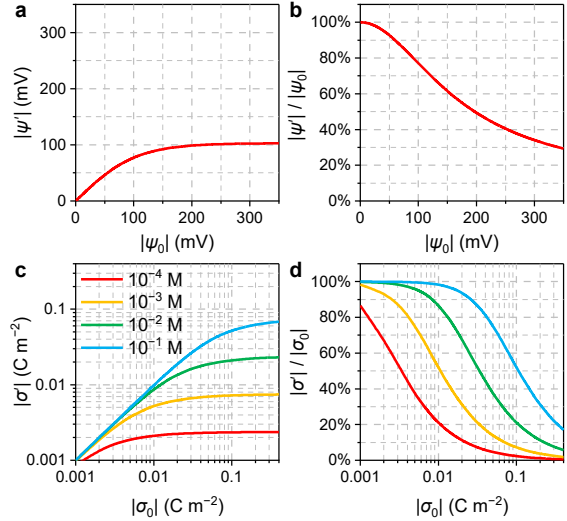

Fig. 3 The results of (**a**) $|\psi'|$ and (**b**) the ratios of $|\psi'|/|\psi_0|$ vary with $|\psi_0|$, and (**c**) $|\sigma'|$ and (**d**) the ratios of $|\sigma'|/|\sigma_0|$ vary with $|\sigma_0|$ at different $c$.

(3) The measured $F_{vdW}$ and the fitted $H$ vary greatly.

The experiments conducted by Valmacco et al. [38] clearly show that the existence of surface roughness can significantly decrease the measured $F_{vdW}$ and the fitted $H$. As has also been reviewed in detail by Valmacco et al. [38], substantial disparities between experimental and theoretical $F_{vdW}$ (and the correspondingly fitted $H$) exist in the literature. Taking silica as an example, the measured $F_{vdW}$ and the fitted $H$ reported for the same experimental systems show huge discrepancies with each other and with theoretical predictions. In some studies, $F_{vdW}$ is even found to be entirely undetected.

The above anomalous results are pervasive but commonly not noticed and questioned in the literature.

We are aware that systematic errors may exist because there are always deviations between the actual experimental situations from the ideal theoretical situations. However, although numerous studies have been conducted in measurements and interpretations of DLVO forces, the existence and effects of systematic errors have rarely been mentioned and considered [2,3,12,13].

The precision of the surface force measurement techniques is very high, commonly with the resolution of pN (piconewton) in force and pm (picometer) in distance [2,3]. However, systematic errors in experiments may still bring about unaccountable or erroneous interpretations of experimental results of surface force measurements. This is because, in general, when fitting the experimental force curves with the theory, the interacting surfaces are almost all regarded as absolutely ideal smooth surfaces with absolutely ideal regular geometric shapes (ideal plane, sphere, cylinder, cone, etc.). But the actual situations will certainly not exactly match the ideal situations, to some extent.

The immediate effects of systematic errors are that they can lead to inaccurate measurements and interpretations of DLVO forces, including the measured $F_{EDL}/F_{vdW}$ and the correspondingly fitted $\psi/\sigma/H$. In fact, the effects of systematic errors have been partly observed and investigated in the literature. For example, surface roughness is an obvious source of systematic errors and has been proved to significantly reduce the DLVO forces and the fitted results [20,35,38–41]. The effects of systematic errors from surface geometry (inaccurately obtained AFM tip geometry parameters, including radius and half angle) on AFM tip–surface interaction forces have also been partially analyzed in our previous work [33]. Nevertheless, these studies are very limited and do not provide comprehensive assessments and effective solutions.

Obviously, the systematic errors in experimental systems may directly cause the above pervasive anomalous results. This will be thoroughly discussed in section 4.

In addition, as the foundation for colloid and interface science, the understandings and usages of the EDL theory are still extensively debated [42,43]. Here we must emphasize that because the 0 distance ($D = 0$) between two adjacent charged surfaces of the measured and calculated $F_{EDL}$–$D$ curves normally starts from each charged surface ($x = 0$) but not the origin plane of the diffuse layer ($x = \delta$), the potential/charge density obtained by directly fitting the measured $F_{EDL}$–$D$ curves, $\psi/\sigma$, is the apparent diffuse layer potential/charge density, not the (actual) diffuse layer potential/charge density, $\psi_\delta/\sigma_\delta$. However, most researchers mistakenly believe that $\psi$ is $\psi_\delta$ and should be close or equal to $\psi_\zeta$ [2,12,14,42], and accordingly, $\sigma$ is $\sigma_\delta$ [2,12,14,30]. This will be thoroughly discussed in section 2.

Therefore, based on rigorous and comprehensive arguments, we can reasonably speculate that, the pervasive anomalous results are existing but not noticed and questioned in the literature owing to the two important aspects: (1) the pervasive unreasonable understandings of EDL theory (see section 2 for detail) and (2) the commonly neglected systematic errors (see section 4 for detail). This will be thoroughly discussed in section 5.

## 1.3 Outline of this review

In this review, we aim to point out the pervasive anomalous results and provide corresponding reasonable explanations in measurements and interpretations of DLVO forces. The outline is as follows.

In section 2, we have clarified the basic understandings of EDL theory in measurements and interpretations of DLVO forces. We have mainly emphasized the definition of the fitted potential/charge density ($\psi/\sigma$) and pointed out the pervasive unreasonable understandings of EDL theory throughout the literature.

In section 3, we have provided the basic methods of measurements and interpretations of DLVO forces, including the surface force measurement techniques and their interaction models and the corresponding calculations of DLVO forces.

In section 4, we have identified the obvious but commonly neglected systematic errors and assessed their effects on the measurements and interpretations of DLVO forces. We have also explained why systematic errors are commonly neglected.

In section 5, based on the arguments about EDL theory and systematic errors, we have reasonably-explained the pervasive anomalous results in the literature and further speculated that, the pervasive anomalous results are existing but not noticed and questioned owing to the two important aspects: (1) the pervasive unreasonable understandings of EDL theory and (2) the commonly neglected systematic errors.

In section 6, based on the characteristics of the EDL theory and systematic errors, we have further constructed theoretical and experimental methods and provided experimental verifications for correcting systematic errors in measurements and interpretations of DLVO forces.

Finally, in section 7, we call for careful consideration of the EDL theory and systematic errors in measurements and interpretations of DLVO forces. On these bases, we can interpret the experimental results properly and promote the development of EDL theory, colloid and interface science, and many related fields.

## 2  Clarifications on the EDL theory

On the interpretations of $F_{EDL}$ in DLVO forces, correctly describing the EDL theory is the first step. In section 2, based on the EDL theory, we highlight that, because the 0 distance ($D = 0$) between two adjacent charged surfaces of the measured and calculated $F_{EDL}$–$D$ curves normally starts from each charged surface ($x = 0$) but not the origin plane of the diffuse layer ($x = \delta$), the potential/charge density obtained by directly fitting the measured $F_{EDL}$–$D$ curves, $\psi/\sigma$, is the apparent diffuse layer potential/charge density, not the (actual) diffuse layer potential/charge density, $\psi_\delta/\sigma_\delta$, nor the surface potential/charge density, $\psi_s/\sigma_s$ (section 2.1). And quantitatively, we have $\psi \gg \psi_\delta$ and $\sigma \gg \sigma_\delta$ (section 2.2). Besides, we have emphasized the difference between $\psi/\sigma$ and $\psi_s/\sigma_s$, the problems about $\psi_\zeta$, the limits of $\psi_\delta/\sigma_\delta$, and the characteristics of $F_{EDL}$ (sections 2.3–2.6). Then, we point out the pervasive unreasonable understandings of EDL theory throughout the literature, by showing that most researchers mistakenly believe that $\psi$ is $\psi_\delta$ and should be close or equal to $\psi_\zeta$, and accordingly, $\sigma$ is $\sigma_\delta$ (section 2.7). The details are as follows.

### 2.1  What exactly is the definition of the fitted potential/charge density ($\psi/\sigma$)?

According to the EDL theory, a diffuse layer is formed at a charged solid–liquid interface. A basic assumption of the Gouy–Chapman (GC) model is that the counterions are mobile and only form non-specific adsorption on the charged surface (all obey the Poisson–Boltzmann (PB) equation). While due to the finite size and possible hydration of ions, other models (e.g., the Stern layer model or the triple-layer model (TLM)) suggest that an immobile and tight adsorption layer is formed adjacent to the charged surface due to specific adsorption, which is called Stern layer or Helmholtz layer (the counterions in which disobey the PB equation) [1–3,30,44]. Therefore, the origin plane of the diffuse layer (also called GC layer) is located at a certain distance $x = \delta$ from the charged surface, which is also called the Stern plane or the outer Helmholtz plane (OHP). **The potential/charge density at the origin plane of the diffuse layer, also called the diffuse layer potential/charge density, can be donated by $\psi_\delta/\sigma_\delta$** (Fig. 4a).

However, it is extremely important to highlight that, **because the 0 distance ($D = 0$) between two adjacent charged surfaces of the measured and calculated $F_{EDL}$–$D$ curves normally starts from each charged surface ($x = 0$) but not the origin plane of the diffuse layer ($x = \delta$)** (Fig. 4b) [2,3], **the potential obtained by directly fitting the measured $F_{EDL}$–$D$ curves, $\psi$, is actually the apparent diffuse layer potential, which is equivalent to extrapolate $\psi_\delta$ to the surface ($x = 0$) based on GC theory** (Eq. (2)), **but not the (actual) diffuse layer potential, $\psi_\delta$, nor the surface potential, $\psi_s$** (Fig. 4). **The correspondingly obtained charge density, $\sigma$, is the apparent diffuse layer charge density, which is equivalent to extrapolate $\sigma_\delta$ to the surface ($x = 0$), but not the (actual) diffuse layer charge density, $\sigma_\delta$, nor the surface charge density, $\sigma_s$.**

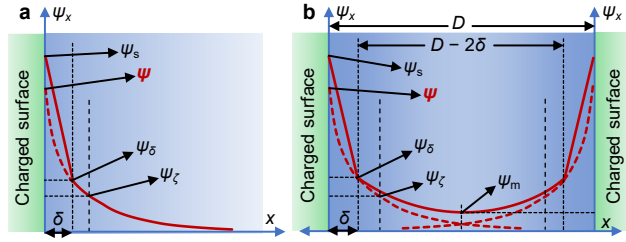

Fig. 4 Schematic descriptions of the EDL of (**a**) an isolated and (**b**) two adjacent charged surfaces.

## 2.2 Difference between $\psi/\sigma$ and $\psi_\delta/\sigma_\delta$

We should seriously consider the parameter $\delta$. As shown in Fig. 5, the ratios of $\psi_x/\psi$ can be obtained by Eqs. (2) and (3), and the ratios of $\sigma_x/\sigma$ can be obtained by substituting Eqs. (2) and (7) into Eq. (4). The results clearly show that the potential/charge density ($\psi_x/\sigma_x$) decays sharply at small $x$, even more drastic than exponential decay.

In practice, $\delta$ is hard to tell. $\delta$ can vary greatly for different counterions and at different $c$ (e.g., around 0.4–1.6 nm) [45,46]. While according to Crothers et al., [30], $\delta \approx r_x + 2r_H$ (around 0.8 nm for alkali metal cations), where $r_x$ is the crystal ionic radius and $r_H$ is the hydrated radius.

By further plotting the ratios of $\psi/\psi_\delta$ and $\sigma/\sigma_\delta$, respectively (Fig. 6, see SI section S1 for calculation detail), we can easily find that **very small $\delta$ (e.g., 0.8 nm) can make $\psi \gg \psi_\delta$ and $\sigma \gg \sigma_\delta$**. More specifically, the $\psi/\sigma$ can become several or even hundreds of times greater than $\psi_\delta/\sigma_\delta$, respectively, especially at larger $\psi_\delta/\sigma_\delta$, $\delta$, and $c$. Therefore, **$\psi/\sigma$ and $\psi_\delta/\sigma_\delta$ should be strictly distinguished**.

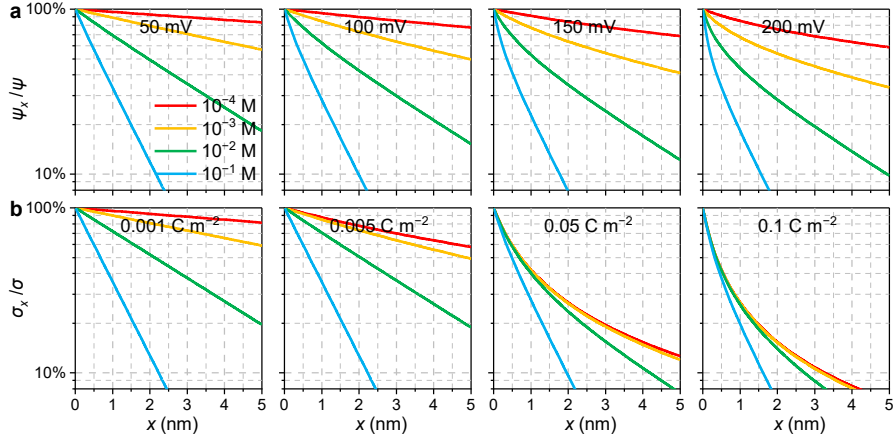

Fig. 5 The ratios of (**a**) $\psi_x/\psi$ and (**b**) $\sigma_x/\sigma$ varying with $x$ at different $\psi$ or $\sigma$ and $c$.

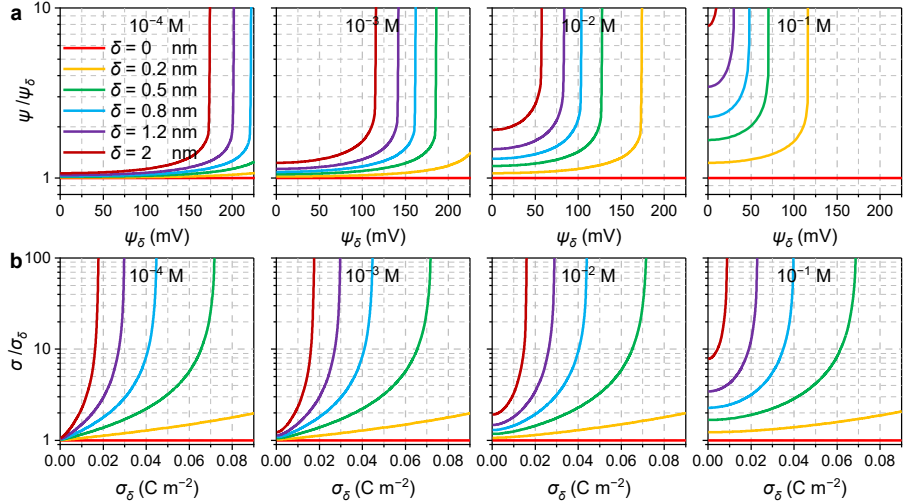

Fig. 6 The ratios of (**a**) $\psi/\psi_\delta$ varying with $\psi_\delta$ and (**b**) $\sigma/\sigma_\delta$ varying with $\sigma_\delta$ at different $\delta$ and $c$.

In addition, it seems possible to convert between $\psi_\delta/\sigma_\delta$ and $\psi/\sigma$ based on GC theory (Eq. (2)) by considering $\delta$. However, this requires very precise knowledge of $\delta$, which is normally difficult or even impossible. In this respect, **$\psi_\delta/\sigma_\delta$ can never be obtained by measurements and interpretations of DLVO forces unless it is extrapolated from the fitted $\psi/\sigma$ after $\delta$ is known**. In fact, even though $\psi/\sigma$ is not the real potential/charge density at any position (by its definition), it is sufficient for us to obtain $\psi/\sigma$ in measurements and interpretations of DLVO forces. $\delta$ and $\psi_\delta/\sigma_\delta$ are only required to be obtained when exploring the quantitative characteristics of the Stern layer.

It should also be noted that, the above clarifications imply that the difference between applying $D - 2\delta$ instead of $D$ is significant (Fig. 4b), even $\delta$ is very small. Therefore, the accurate determination of $D$ is extremely important in surface force measurements. We can readily realize that nanoscale surface roughness may strongly affect the accurate determination of $D$ and further affect the measurements and interpretations of DLVO forces. In fact, the effects of surface roughness can be astonishing, as extensively demonstrated in sections 4, 5, and 6.

### 2.3 Difference between $\psi/\sigma$ and $\psi_s/\sigma_s$

It's worth noting that, for **strongly hydrated counterions** (e.g., $Li^+$ and $Na^+$) which are not easy to adsorb specifically to the charged surfaces, the GC model is generally regarded as valid and we have $\psi/\sigma \approx \psi_s/\sigma_s$ (Fig. 4a) [44,47]. In contrast, when specific adsorption is present, $\psi/\sigma$ and $\psi_s/\sigma_s$ may differ considerably.

### 2.4 Disscussion on $\psi_\zeta$ (zeta potential)

Due to the convenience of measurements, the zeta potential, $\psi_\zeta$, is commonly used and has become an important indicator for characterizing EDL properties.

$\psi_\zeta$ is assumed to arise on the hypothetical slip plane or shear plane ($x = d_\zeta$) (Fig. 4a). It is generally supposed that the slip/shear plane is located close to the origin plane of the diffuse layer **($d_\zeta \geq \delta$),** which means $\psi_\zeta$ is always smaller than or close to $\psi_\delta$ [44,48,49]. While theoretical simulations reveal that $d_\zeta \approx 15-20$ Å [43], which seems to be much larger than $\delta$. This implies that $\psi_\zeta$ may be much smaller than, rather than close to, $\psi_\delta$. But in either case, we have **$\psi \gg \psi_\zeta$**. And likewise, **$\psi$ and $\psi_\zeta$ should also be strictly distinguished**.

In addition, many problems about the definition and determination of $\psi_\zeta$ remain to be seriously considered.

Similar to $\delta$, the precise $d_\zeta$ is normally difficult or even impossible to know. And $d_\zeta$ may also vary as greatly as $\delta$ for different counterions and at different $c$. Therefore, **it is not possible to predict $\psi/\sigma$ from the measured $\psi_\zeta$ unless $d_\zeta$ is known**.

Besides, the requirement to obtain $\psi_\zeta$ accurately is that electrokinetic theories are correctly applied within their range of validity. This is often difficult to ensure or even completely ignored, leading to wide variations between the results of different researchers [49]. Thus, the measured $\psi_\zeta$ usually appears to be unreliable. For example, according to section 2.5, $\psi_\delta$ has limits. If we assume that $\psi_\zeta$ should be smaller than or close to $\psi_\delta$, then $\psi_\zeta$ should not exceed the limits of $\psi_\delta$ either. However, substantial results of measured $\psi_\zeta$ in the literature (refer to the references cited in this section) can exceed the allowable range. Moreover, the frequently found results of non-monotonically varying $\psi_\zeta$ with $c$ are clearly at odds with the EDL theory [50–52].

As a result, a few researchers (especially early on) still call for caution in the use of $\psi_\zeta$ [1,43,44,49,53]. This is well worth thinking about.

But as can be seen in the literature, a large number of researchers have assumed that $\psi_\zeta$ is approximate or equivalent to $\psi_\delta$ and boldly applied it in their studies [15,30,46,54,55]. What's worse, by mistaking $\psi_\delta$ for $\psi$, most researchers have accepted that $\psi_\zeta$ should be close or equal to $\psi$ [2,12,14,42]. This will be further discussed in section 2.7.

In summary, it is important to highlight that: (1) $\psi$ and $\psi_\zeta$ should be strictly distinguished; (2) it is risky to approximate $\psi_\zeta$ as $\psi_\delta$; and (3) $\psi_\zeta$ should be applied with caution in related research fields due to the problems about its definition and determination.

## 2.5 Limits of $\psi_\delta/\sigma_\delta$

As also shown in Fig. 6, $\psi_\delta/\sigma_\delta$ has limits when $\psi/\sigma$ is large. This is simply because $|\psi_x|/|\sigma_x|$ ($x \neq 0$, Eqs. (2) and (4) in section 3.2) has an important characteristic that it tends to be maximum and constant and independent of $\psi/\sigma$ when $\psi/\sigma$ is large. This can be directly reflected by the existence of the tanh($x$) function ($t$, Eq. (3)), whose absolute value tends to 1 with large $\psi/\sigma$.

The limits of $|\psi_\delta|/|\sigma_\delta|$ varying with $c$ at different $\delta$ when $\psi/\sigma$ tends to infinity ($\psi/\sigma \to \infty$ and $|t| \to 1$) are illustrated in Fig. 7. For example, assuming $\delta = 0.8$ nm, $|\psi_\delta|$ has the maximum values of 223, 164, 105, 48.4, and 7.6 mV and $|\sigma_\delta|$ has the maximum values of 0.046, 0.046, 0.045, 0.041, and 0.018 C m$^{-2}$ at representative $c$ of $10^{-4}$, $10^{-3}$, $10^{-2}$, $10^{-1}$ and 1 M, respectively.

However, the reality is that substantial results of fitted $\psi/\sigma$ can exceed the allowable range of $\psi_\delta/\sigma_\delta$ (refer to sections 2.6 and 1.2). This again demonstrates that the fitted potential/charge density is $\psi/\sigma$ rather than $\psi_\delta/\sigma_\delta$, and $\psi/\sigma$ and $\psi_\delta/\sigma_\delta$ should be strictly distinguished (sections 2.1 and 2.2).

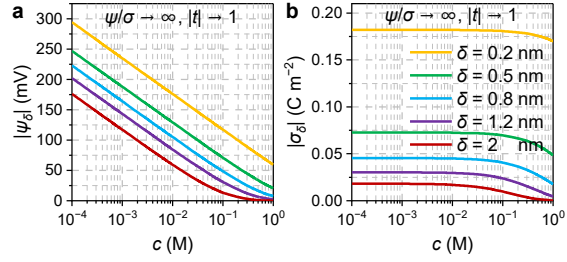

Fig. 7 The limits of (**a**) $|\psi_\delta|$ and (**b**) $|\sigma_\delta|$ varying with $c$ at different $\delta$ when $\psi/\sigma$ tends to infinity ($\psi/\sigma \to \infty$ and $|t| \to 1$).

## 2.6 Characteristics of $F_{EDL}$

Similar to section 2.5, **$|F_{EDL}|$ also has the important characteristic that it tends to be maximum and constant and independent of $\psi/\sigma$ when $\psi/\sigma$ is large**, according to the calculation formulas of $F_{EDL}$ (e.g., LSA (Eqs. (6) and (9)) and exact calculation methods [1,2]). Therefore, as long as $\psi/\sigma$ can be guaranteed to be large enough, $|F_{EDL}|$ can be calculated regardless of the exact value of $\psi/\sigma$. By plotting the $F_{EDL}$ curves varying with $\psi/\sigma$ at different $c$ (Fig. 8), we can make a rough estimation that $|F_{EDL}|$ can be approximately regarded to be maximum and constant at **around $|\psi| \geq 150$ mV or $|\sigma| \geq 0.05$ C m$^{-2}$**, and this estimation is more accurate at higher $c$ for $\psi$ or lower $c$ for $\sigma$.

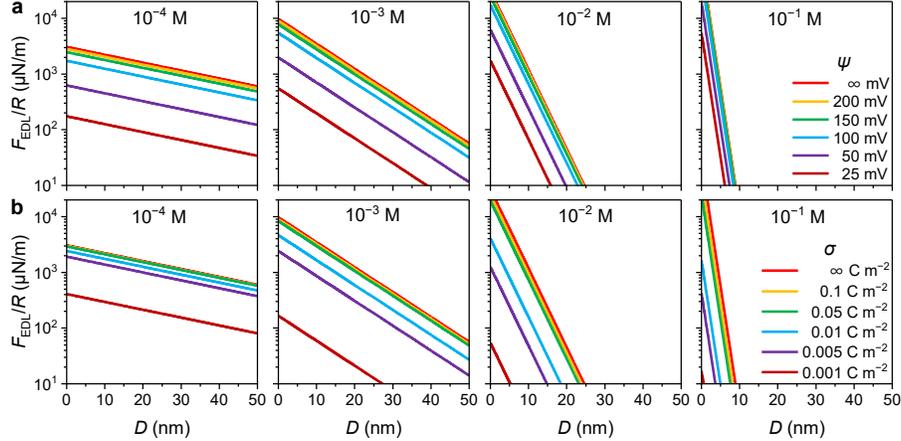

Fig. 8 The $F_{EDL}$ curves varying with (**a**) $\psi$ and (**b**) $\sigma$ at different $c$ calculated by Eq. (6).

In general, the $\sigma_s$ of commonly used materials can be sufficiently large (some examples are shown in Fig. 9). The basal plane of muscovite mica [K(Si$_3$Al)(Al$_2$)O$_{10}$(OH)$_2$] mainly used in SFA has a constant $\sigma_s$ of $-0.34$ C m$^{-2}$ (1/0.47 = 2.1 e nm$^{-2}$), which is calculated from the crystal lattice parameters and supported by experiments based on ion exchange [56,57]. As for silica usually used in AFM, the $\sigma_s$ can reach −0.05 C m$^{-2}$ at pH ~7.5 and |$\sigma_s$| increases sharply with the increase of pH [46,58–61]. The commonly used polystyrene latex particles functionalized with different groups (e.g., carboxyl, sulfate, and amidine) in AFM also have |$\sigma_s$| ≥ 0.05 C m$^{-2}$ at appropriate pH [47,62]. For silicon nitride (Si$_3$N$_4$) commonly used as an AFM tip material, the $\sigma_s$ can reach −0.05 C m$^{-2}$ at pH ~9 and |$\sigma_s$| increases sharply with the increase of pH [63,64].

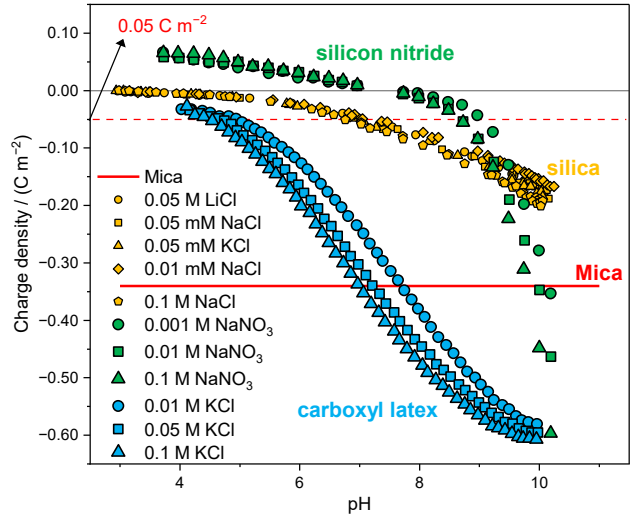

Fig. 9 The $\sigma_s$ of commonly used materials measured by potentiometric titration. Data sources: Silica [45,46], silicon nitride [64], and carboxyl latex [62].

As clarified in section 2.3, for strongly hydrated counterions (e.g., Li$^+$ and Na$^+$) which are not easy to adsorb specifically to the charged surfaces, $\psi/\sigma$ should be close to $\psi_s/\sigma_s$. Therefore, for the commonly used materials, the $\psi/\sigma$ range making |$F_{EDL}$| tend to be maximum and constant can be easily satisfied at appropriate $c$ and pH.

In fact, the characteristic that |$F_{EDL}$| tends to be maximum and constant can be frequently observed in the experimental results. For example, the interaction forces between mica measured at near neutral to basic pH (5.7–10) in 10$^{-2}$ or 10$^{-3}$ M Li$^+$ solutions are almost exactly the same, as also pointed out by the authors [15]. While according to interaction forces measured between silica [19–21], the fitted $\psi$ are basically the same at different pH (around >7, at same $c$), which obviously indicates the same $F_{EDL}$ (the reasons for small fitted $\psi$ are explained in section 5.1).

Obviously, the fitted $\psi/\sigma$ also has limits. **The exact value of $\psi/\sigma$ cannot be obtained when $\psi/\sigma$ is large enough** (around |$\psi$| ≥ 150 mV or |$\sigma$| ≥ 0.05 C m$^{-2}$).

## 2.7 Pervasive unreasonable understandings of EDL theory

Based on the above descriptions, we have obtained some basic understandings of EDL theory. Most importantly, we have $\psi \gg \psi_\delta \geq \psi_\zeta$ and $\sigma \gg \sigma_\delta$. However, despite the above clarifications, similar basic understandings of EDL theory can hardly be found in the literature. On the contrary, **pervasive unreasonable understandings of EDL theory can be found everywhere: most researchers mistakenly believe that $\psi$ is $\psi_\delta$ and should be close or equal to $\psi_\zeta$ [2,12,14,42], and accordingly, $\sigma$ is $\sigma_\delta$ [2,12,14,30].**

The theoretical assumptions of the unreasonable understandings of EDL theory are often specious. In the measurements and interpretations of DLVO forces, it is usually taken for granted that, since $F_{EDL}$ originates from the overlap of the diffuse layers, the potential/charge density obtained by directly fitting the measured $F_{EDL}$–$D$ curves ($\psi/\sigma$) should naturally be the diffuse layer potential/charge density ($\psi_\delta/\sigma_\delta$). This is the key point. In fact, we can easily recognize the difference between $\psi/\sigma$ and $\psi_\delta/\sigma_\delta$, which differ in position by $\delta$ (Fig. 4). However, until one sees the calculation results (Fig. 6), it may be difficult to imagine that even a very small $\delta$ (e.g., only 0.8 nm) can cause huge differences between $\psi/\sigma$ and $\psi_\delta/\sigma_\delta$. The case of $\psi_\zeta$ is similar.

What's worse, researchers even believe that the surface charge is completely neutralized (or eliminated) by specifically adsorbed counterions adjacent to the charged surface (before the origin plane of the diffuse layer), and the remaining charge corresponding to the diffuse layer is the one that is reflected by surface force measurements, the proportion of which can even be smaller than 1% [2,12,14,30].

In fact, such unreasonable understandings are unreliable and problematic in two respects:

(1) It is not only the remaining charge corresponding to the diffuse layer that is reflected by surface force measurements.

As clearly illustrated in sections 2.1 and 2.2, because the 0 distance ($D = 0$) between two adjacent charged surfaces of the measured and calculated $F_{EDL}$–$D$ curves normally starts from each charged surface ($x = 0$) but not the origin plane of the diffuse layer ($x = \delta$) (Fig. 4b), even assuming that the counterions adjacent to the surface completely neutralize (eliminate) the surface charge (chemical bonds are formed), the effects of size of the dehydrated and hydrated counterions (the parameters $\delta$ and $d_\zeta$), which make $\psi/\sigma \gg \psi_\delta/\sigma_\delta$ and $\psi \gg \psi_\zeta$, cannot be neglected. This obviously indicates that it is not only the remaining charge corresponding to the diffuse layer that is reflected by surface force measurements. The fitted charge density ($\sigma$) must always be much greater than the diffuse layer charge density ($\sigma_\delta$), unless we make the 0 distance ($D = 0$) of the measured $F_{EDL}$–$D$ curves start from the origin plane of the diffuse layer ($x = \delta$).

(2) The viewpoint that the surface charge is completely neutralized (or eliminated) by specifically adsorbed counterions adjacent to the charged surface may be misleading.

Assuming that we choose a plane at any position in the diffuse layer as the object of study, it is true that the charge density of this plane is defined only by the remaining charge beyond this plane. However, it is definitely not appropriate to assume that the counterions before this plane have completely neutralized (eliminated) the surface charge, which is still contributing to strong electrical field and EDL (electrostatic) force! By definition, only chemical bonding can truly be considered to completely neutralize (eliminate) the surface charge (i.e., no longer contribute to the electrical field and EDL (electrostatic) force). In contrast, counterions act as a screening effect on the surface charge (electrical field). The specifically adsorbed counterions normally have a stronger screening effect than electrostatically adsorbed counterions on the electrical field generated by the surface charge but cannot completely neutralize (eliminate) the surface charge.

What's more, not all counterions form specific adsorption. Due to the presence of a hydrated layer, the strongly hydrated counterions (e.g., $Li^+$ and $Na^+$) generally only form non-specific adsorption on the charged surface (section 2.3).

In addition, the unreasonable understandings are also usually attributed to $H^+$. It has been suggested that the $K^+$ initially held on the mica surface should be almost completely exchanged for $H_3O^+$ in distilled water at pH 6.5–7.0 [57,65]. However, the interaction forces measured at near neutral to basic pH (5.7–10) in $10^{-2}$ or $10^{-3}$ M $Li^+$ solutions are almost exactly the same [15], which obviously indicate that the influence of $H^+$ can be ignored at least under this pH range. The $H^+/K^+$ exchange experiment on montmorillonite can also clearly prove that $H^+$ cannot completely replace $K^+$ even at very lower pH (= 2, 0.01 M $H^+$) [66].

In summary, the pervasive unreasonable understandings of EDL theory should be seriously considered. Moreover, pervasive anomalous results in the measurements and interpretations of DLVO forces can also be found everywhere. This has been mentioned in section 1.2 and will be further discussed and explained in section 5.

## 3 Basic methods of measurements and interpretations of DLVO forces

In section 3, based on the clarifications on the EDL theory (section 2), the basic methods of measurements and interpretations of DLVO forces, including the surface force measurement techniques and their interaction models (section 3.1) and the corresponding calculations of DLVO forces (section 3.2), are described as follows.

### 3.1 Techniques and interaction models

At present, there are three main surface force measurement techniques, namely SFA, AFM microsphere (colloidal) probe technique, and AFM nanoscale tip technique, each of which has different geometry features for the interaction models (Fig. 1d–g) [2,3]:

(1) SFA

Developed by Tabor and Winterton [5], Israelachvili and Tabor [6], and Israelachvili and Adams [4,7,8], SFA generally requires two macroscopic cylindrical surfaces, and the interaction model is the two-macroscopic-cylinder interaction model (denoted by c–c, Fig. 1d). Mica is an ideal material suitable for SFA due to its atomically smooth macroscopic surface.

(2) AFM microsphere (colloidal) probe technique

The AFM was invented by Binnig et al. [9] in 1986. In order to fit the theory to the experimental results, a microsphere (colloidal particle) with regular spherical shape is usually attached to the AFM probe cantilever, and the interaction models can be simplified to the microsphere–flat surface interaction model (denoted by s–f) and two-microsphere interaction model (denoted by s–s), respectively (Fig. 1e–f). Due to the requirement of spherical shape, generally only artificial materials like silica, polystyrene latex, and other few materials can be used.

(3) AFM nanoscale tip technique

AFM nanoscale tip technique can overcome the limitation of sample geometry relatively well due to the nanoscale AFM tip ends. The interaction model can be simplified to the conical tip with flat circular end–flat surface interaction model (denoted by cf–f) (Fig. 1g). The cf–f model is recommended for more accuracy among the various AFM tip–surface interaction models, and the reasons have been explained in our previous work [33]. The tip material is usually $Si_3N_4$ or Si ($SiO_2$), and the flat surface materials can be very broad, represented by silica, clay minerals, metal oxides, and others.

## 3.2 Calculations of DLVO forces

Since the 0 distance ($D = 0$) of the measured and calculated $F_{EDL}$–$D$ curves normally starts from each charged surface ($x = 0$) (section 2), all equations below are uniformly derived with reference to the charged surface ($x = 0$) for convenience and consistency with the literature.

The **potential at distance $x$ from a charged surface, $\psi_x$,** in a 1:1 electrolyte solution satisfies the Poisson–Boltzmann (PB) equation [1,2]

$$\frac{d^2\psi_x}{dx^2} = \frac{qn_{i0}}{\varepsilon\varepsilon_0}\sinh\left(\frac{q\psi_x}{k_BT}\right) \tag{1}$$

where $x$ is the distance from a charged surface, $q$ is the elementary charge ($1.6022 \times 10^{-19}$ C), $n_{i0}$ is the bulk number concentration of ion $i$ in the electrolyte solution, $\varepsilon$ is the relative permittivity (80.1 for water), $\varepsilon_0$ is the dielectric permittivity of vacuum ($8.854 \times 10^{-12}$ C$^2$ J$^{-1}$ m$^{-1}$), $k_B$ is the Boltzmann constant ($1.3806 \times 10^{-23}$ J K$^{-1}$), and $T$ is the absolute temperature (298K for room temperature).

The Gouy–Chapman (GC) theory gives the exact solution of PB equation in a 1:1 electrolyte solution [1,2]

$$\psi_x = \frac{4k_BT}{q}\operatorname{atanh}\left(t e^{-\kappa x}\right) \tag{2}$$

and

$$t = \tanh\left(\frac{q\psi}{4k_BT}\right) \tag{3}$$

where e is the base of the natural logarithm, $\kappa^{-1}$ is the Debye length and $\kappa = \sqrt{\dfrac{2q^2 N_A c}{\varepsilon\varepsilon_0 k_B T}}$, $N_A$ is the Avogadro constant ($6.023 \times 10^{23}$ mol$^{-1}$), $c$ is the molar concentration of the electrolyte solution, and $\psi$ is the potential obtained by directly fitting the measured $F_{EDL}$–$D$ curves, and also the apparent diffuse layer potential, which is equivalent to extrapolate $\psi_\delta$ to the surface ($x = 0$), as described in section 2.1.

We can obtain the relationship between the corresponding charge density at distance $x$ from a charged surface, $\sigma_x$, and $\psi_x$ based on the Grahame equation [2,67]:

$$\sigma_x = \frac{2\kappa\varepsilon\varepsilon_0 k_B T}{q}\sinh\left(\frac{q\psi_x}{2k_BT}\right) \tag{4}$$

As mentioned in section 1, the DLVO forces ($F_{DLVO}$) can be expressed by

$$F_{DLVO} = F_{EDL} + F_{vdW} \qquad (5)$$

$F_{EDL}$ and $F_{vdW}$ for c–c, s–f, and s–s models

For c–c, s–f, and s–s models (Fig. 1d–f), $F_{EDL}$ for two adjacent charged surfaces with isolated $\psi$, $\psi_1$ and $\psi_2$, in a 1:1 electrolyte solution has the well-known approximate expression, linear superposition approximation (LSA) [2,8,33,68]:

$$F_{EDL} = At_1t_2\kappa R e^{-\kappa D} \Leftrightarrow \log F_{EDL} = -\kappa D \log e + \log(At_1t_2\kappa R)$$
$$\Leftrightarrow \ln F_{EDL} = -\kappa D + \ln(At_1t_2\kappa R) \qquad (6)$$

where A is a self-defined constant given by $A = \dfrac{64\pi\varepsilon\varepsilon_0 k_B^2 T^2}{q^2}$ (9.40 × 10$^{-11}$ N), π is ratio of the circumference of a circle to its diameter, $t_1t_2$ is calculated by substituting $\psi_1$ and $\psi_2$ into Eq (3), R is the equivalent radius, D is the distance between two charged surfaces, log is the logarithm to any base, and ln is the natural logarithm. For s–f model, R is the same as the sphere radius; for s–s model, $R = \dfrac{R_1 R_2}{R_1 + R_2}$; and for c–c model, $R = \sqrt{R_1 R_2}$ (see Fig. 1d–f).

Eq. (6) implies that **$F_{EDL}$ is linearly related to D in logarithmic y-axis (to any base). Especially, the slope is −κ in natural logarithmic y-axis**.

According to the Grahame equation (Eq. (4)), we have the relationship between the corresponding σ and ψ (set x = 0, i.e., replace $\psi_x$ by ψ):

$$\sigma = \frac{2\kappa\varepsilon\varepsilon_0 k_B T}{q} \sinh\left(\frac{q\psi}{2k_B T}\right) \qquad (7)$$

The only approximation condition for the LSA expression (Eq. (6)) is to assume that the midplane potential $\psi_m$ (not ψ) is extremely small (Fig. 4b, $|\psi_m|$ < 25 mV, $\left|\dfrac{q\psi_m}{k_B T}\right| \ll 1$), which can be easily satisfied at around $D > \kappa^{-1}$. So the LSA expression agrees well with the exact calculation results at $D > \kappa^{-1}$, as pointed out in the literature [2,8] and our previous work [33]. Therefore, the LSA expression can be used instead of the exact calculation methods to fit the $D > \kappa^{-1}$ part of the $F_{EDL}$–$D$ curves, and all the correct fitted results of $\psi/\sigma$ for the isolated charged surfaces can be obtained.

$F_{vdW}$ can be written as an explicit function of $D$ (ignoring retardation effects) [2,3]:

$$F_{vdW} = -\frac{HR}{6D^2} \Leftrightarrow \log(-F_{vdW}) = -2\log D + \log\frac{HR}{6} \tag{8}$$

where $H$ is the Hamaker constant.

$F_{vdW}$ generally have much shorter force range (a few nm) than $F_{EDL}$.

Eq. (8) implies that **$-F_{vdW}$ is linearly related to $D$ with a slope of $-2$ in logarithmic x$y$-axis**.

$F_{EDL}$ and $F_{vdW}$ for cf–f model

Similarly, for cf–f model (Fig. 1g), the $F_{EDL}$ and $F_{vdW}$ can be expressed as [33]

$$\begin{aligned} F_{EDL} &= At_1t_2\left(\frac{1}{2}\kappa^2 R^2 + \tan^2\theta + \kappa R\tan\theta\right)e^{-\kappa D} \\ &\Leftrightarrow \log F_{EDL} = -\kappa D\log e + \log\left[At_1t_2\left(\frac{1}{2}\kappa^2 R^2 + \tan^2\theta + \kappa R\tan\theta\right)\right] \\ &\Leftrightarrow \ln F_{EDL} = -\kappa D + \ln\left[At_1t_2\left(\frac{1}{2}\kappa^2 R^2 + \tan^2\theta + \kappa R\tan\theta\right)\right] \end{aligned} \tag{9}$$

and

$$F_{vdW} = -\frac{H}{6}\left(\frac{\tan^2\theta}{D} + \frac{R\tan\theta}{D^2} + \frac{R^2}{D^3}\right) \tag{10}$$

where $R$ is the equivalent tip end radius and $\theta$ is the equivalent tip cone half angle.

Eq. (9) implies that **$F_{EDL}$ is linearly related to $D$ in logarithmic $y$-axis (the slope is $-\kappa$ in natural logarithmic $y$-axis)**.

It is worth noting that, **$-F_{vdW}$ is not exactly linearly related to $D$ in logarithmic x$y$-axis**, as shown in Eq. (10). But **at $D \ll R$**, Eq. (10) can be approximated as $F_{vdW} = -\frac{HR^2}{6D^3} \Leftrightarrow \log(-F_{vdW}) = -3\log D + \log\frac{HR^2}{6}$, which is the same as the expression of flat surface–flat surface interaction model (denoted by f–f, also showed in our previous work [33]), and implies that $-F_{vdW}$ is approximately linearly related to $D$ with a slope of $-3$ in logarithmic x$y$-axis.

## 4 Identification and effects of systematic errors

In section 4, with careful inspection, the obvious but commonly neglected systematic errors mainly from three possible sources, surface roughness, surface geometry, and spring constant, can be identified in measurements and interpretations of DLVO forces (section 4.1). Through concise and simple theoretical calculations, we have clearly demonstrated that the identified systematic errors are entirely sufficient to cause the results of measurements and interpretations of DLVO forces to deviate substantially, or even completely, from reality, not just the problem of accuracy (section 4.2). Finally, we try to explain why systematic errors are commonly neglected, which can be attributed to (1) the commonly neglected implicit characteristics of the effects of systematic errors, (2) the commonly neglected strong amplifying effects on systematic errors, and (3) the inappropriate and incomplete understandings of effects of systematic errors (section 4.3). The details are depicted as follows.

## 4.1 Identification of systematic errors

As mentioned in section 1.2, systematic errors may exist because there are always deviations between the actual experimental situations from the ideal theoretical situations.

With careful inspection, **the obvious but commonly neglected systematic errors mainly from three possible sources, surface geometry, surface roughness, and spring constant**, can be identified in measurements and interpretations of DLVO forces (Fig. 10). Being systematic errors implies that their magnitude and effects are exactly the same in the same batch of experiments.

The details are depicted as follows.

### 4.1.1 Surface geometry

DLVO forces are directly related to the surface geometry, which means systematic errors of surface geometry parameters (Fig. 1d–g and section 3.2) can directly affect the calculations of DLVO forces. Because the maximum $D$ range of the DLVO forces can generally only reach tens to hundreds of nm (nanometers), while the used materials are normally μm (micrometer)- to cm (centimeter)-sized (for AFM nanoscale tips with nanoscale tip ends (Fig. 1g), they still have μm-sized tip height), only small parts of the surfaces participate in the interaction forces. Therefore, the deviation of the effective (actual) surface geometry parameters from the apparent (ideal) ones is mainly originated from two aspects: (1) the difference between the local and the global geometry and (2) the accuracy of the measurements.

The surface geometry parameters are generally determined by scanning electron microscopy (SEM) or even optical microscopy, which assumes that the surfaces have ideal regular geometric shapes (ideal plane, sphere, cylinder, cone, etc.). However, the characterization methods are difficult to discern the difference between the local and the global features, and also are easy to introduce large systematic errors [69]. For example, for μm-sized microsphere probes, experimental results indicate that the difference between the local and the global values for the radius of curvature can be significant [69,70]. Theoretical studies have shown that the local surface imperfections of a spherical lens with cm-sized radius can increase or decrease the magnitude of the measured force by several tens of percent when compared with the case of ideal spherical lens [71].

As illustrated in Fig. 10a, for SFA and AFM microsphere probe technique, the local small part of an interacting surface may differ from the global surface, which can introduce systematic errors of $R$. Due to the systematic errors of $R$, the apparent $R$ (denoted by $R'$) used in calculations may change $E_R$ times relative to the effective (actual) $R$ (denoted by $R_0$), and we have $R' = E_R R_0$. $E_R$ is a self-defined systematic error parameter called error coefficient of $R$ and $E_R - 1$ is the relative error of $R$. For SFA, $E_R - 1$ is commonly reported in the order of 10%–20% [72,73]. For AFM microsphere probe technique, $E_R - 1$ can reach 20% or more [69,70].

Similar for AFM nanoscale tips, as pointed out in our previous work [33], it is easy to notice that there may be large systematic errors (even in the magnitude of several tens of percent) for the two surface geometry parameters, equivalent tip end radius $R$ and equivalent tip cone half angle $\theta$ (Fig. 1g), which usually are directly observed through SEM images [74–76]. Due to the systematic errors of $R$ and $\theta$, the apparent $R$ and $\theta$ (denoted by $R'$ and $\theta'$) used in calculations may change $E_R$ and $E_\theta$ times relative to the effective (actual) $R$ and $\theta$ (denoted by $R_0$ and $\theta_0$), respectively, and we have $R' = E_R R_0$ and $\theta' = E_\theta \theta_0$. $E_R$ and $E_\theta$ are self-defined systematic error parameters called error coefficient of $R$ and error coefficient of $\theta$, and $E_R - 1$ and $E_\theta - 1$ are the relative error of $R$ and $\theta$, respectively.

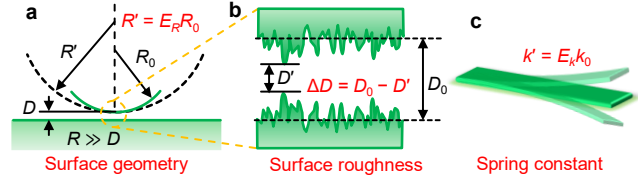

Fig. 10 Schematic description of systematic errors mainly from three possible sources: (**a**) surface geometry, (**b**) surface roughness, and (**c**) spring constant.

### 4.1.2 Surface roughness

As mentioned in section 2.2, nanoscale surface roughness may strongly affect the accurate determination of $D$ and further affect the interpretations of DLVO forces. Surface roughness is an obvious source of systematic errors. Except for cleaved mica, which has negligible roughness due to its atomically smooth surface, most common materials have different degrees of surface roughness. In recent years, more and more researchers have found that the nanoscale roughness of the material surfaces can significantly reduce the DLVO forces and the fitted results [20,35,38–41]. However, how to quantitatively characterize and eliminate the effects of surface roughness when interpreting the experimental force curves remains questionable.

According to the surface topography measured by AFM [39] (also shown in Fig. 17b5–b7), surface roughness can generally be characterized by the surface spikes, as represented in Fig. 10b. This means the two surfaces no longer continue to approach due to the contact of the surface spikes, which may influence the determination of surface distance $D$.

For SFA, the absolute distance between two surfaces is optically measured using multiple beam interferometry (MBI) by interpreting the fringes of equal chromatic order (FECO) patterns [3]. Hence the effect of surface roughness on $D$ can be eliminated by carefully identifying the effective (actual) interacting surfaces.

However, for AFM, the determination of zero distance ($D = 0$) is obtained by the appearance of the hard contact [3]. This means the measured distance of rough surfaces by AFM is between the apparent surfaces with spikes, not the effective (actual) surfaces. Therefore, it can be inferred that the major effect caused by surface roughness is that there is a difference between the effective (actual) $D$ (denoted by $D_0$) and the apparent $D$ (denoted by $D'$). This difference can be represented by $\Delta D$ ($\Delta D = D_0 - D'$) (Fig. 10b), and $\Delta D$ is a self-defined systematic error parameter called distance offset. For commonly used materials, $\Delta D$ is typically a few nm or even tens of nm as can be inferred from surface topography measured by AFM [39].

It is worth noting that, by definition, $\Delta D$ should be close to the peak-to-valley height ($H_{p-v}$) rather than the root-mean-square (RMS) roughness ($\sigma_{RMS}$), and $\sigma_{RMS}$ should be much smaller than $\Delta D$. However, $\sigma_{RMS}$ is commonly used to evaluate and correct the effects of surface roughness by researchers [20,35,41]. This will be further discussed in section 6.3.

### 4.1.3 Spring constant of the spring devices

Spring constant (or stiffness) $k$ is a crucial parameter due to the inevitable use of the spring devices. The effect of systematic error of $k$ is to cause the apparent (measured) $k$ (denoted by $k'$) used in force measurements to become $E_k$ times the effective (actual) $k$ (denoted by $k_0$). Therefore, we have $k' = E_k k_0$, which also means the measured force becomes $E_k$ times its true value (Fig. 10c). $E_k$ is a self-defined systematic error parameter called error coefficient of $k$, and $E_k - 1$ is the relative error of $k$.

For SFA, $k$ is measured by placing small weights on the lower disk holder and measuring the displacement of the spring in a traveling microscope [4]. The $E_k - 1$ for SFA is reported to be about 3% [6] or about 10% [77]. While for AFM, the reported $E_k - 1$ obtained by the two most commonly used methods, thermal noise and Sader methods, are about 5–20% and 10–30%, respectively [78].

## 4.2 Effects of systematic errors

Based on section 4.1, we may still think that the effects of systematic errors from the above three sources are not significant. As can be seen in the literature, the systematic errors are commonly neglected, either intentionally or unintentionally. But can these systematic errors be neglected?

Through concise and simple theoretical calculations, the effects of the identified systematic errors (quantified by $\Delta D$, $E_R$, $E_\theta$, and $E_k$) on the measurements and interpretations of DLVO forces (i.e., the measured $F_{EDL}/F_{vdW}$ and the fitted $\psi/\sigma/H$) can be easily illustrated. These results can clearly show that the identified systematic errors are entirely sufficient to cause the results of measurements and interpretations of DLVO forces to deviate substantially or even completely from reality, not just the problem of accuracy. The main results can be illustrated by Fig. 11 and summarized as:

(1) For c–c, s–f, s–s, and cf–f models, $F_{EDL}$ is linearly related to $D$ (Fig. 11a, Eqs. (6) and (9)), and the effect of $\Delta D$ is equivalent to move the original theoretical $F_{EDL}$ curve along the $x$-axis to the left by $\Delta D$ (Fig. 11b), while the effect of $E_R$, $E_\theta$, and $E_k$ is equivalent to move the original theoretical $F_{EDL}$ curve along the y-axis up/down (Fig. 11c) (in logarithmic y-axis).

(2) For c–c, s–f, s–s, and cf–f models, the existence of $\Delta D$, positive errors of $R$ ($E_R - 1 > 0$) and $\theta$ ($E_\theta - 1 > 0$), and negative errors of $k$ ($E_k - 1 < 0$) can all (significantly) decrease the measured $F_{EDL}$ and the fitted $\psi/\sigma$, and even cause the fitted $\psi/\sigma$ to be extremely small (smaller than 50% or even 10% of the actual value); while the existence of negative errors of $R$ ($E_R - 1 < 0$) and $\theta$ ($E_\theta - 1 < 0$) and positive errors of $k$ ($E_k - 1 > 0$) can all (significantly) increase the measured $F_{EDL}$ and the fitted $\psi/\sigma$, and even cause the fitted $\psi/\sigma$ to exceed the allowable range of calculation (Fig. 11d–f).

(3) For c–c, s–f, and s–s models, $-F_{vdW}$ is linearly related to $D$ (Fig. 11g, Eq. (8)), and the effect of $\Delta D$ is to change the original theoretical $-F_{vdW}$ curve from a straight line to a curve, and the curvature of the curve increases as $\Delta D$ increases (Fig. 11h); while the effect of $E_R$ or $E_k$ is equivalent to move the original theoretical $-F_{vdW}$ curve along the y-axis up/down (Fig. 11i) (in logarithmic xy-axis). The case of cf–f is similar but somewhat complicated and is omitted.

(4) For c–c, s–f, and s–s models, the existence of $\Delta D$, positive errors of $R$ ($E_R - 1 > 0$), and negative errors of $k$ ($E_k - 1 < 0$) can all (significantly) decrease the measured $F_{vdW}$ and the fitted $H$; while the existence of negative errors of $R$ ($E_R - 1 < 0$) and positive errors of $k$ ($E_k - 1 > 0$) can all (significantly) increase the measured $F_{vdW}$ and the fitted $H$ (Fig. 11j–l). The case for cf–f is similar but somewhat complicated and is omitted.

The detailed results are as follows and the corresponding calculations are in SI section S2.

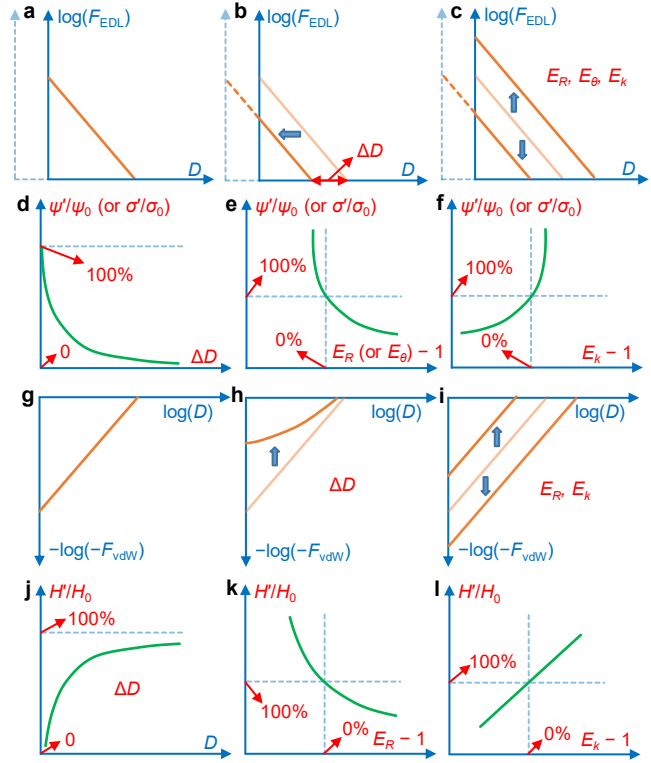

Fig. 11 Schematic description of the effect of systematic errors on the measurements and interpretations of DLVO forces, (**a**–**f**) $F_{EDL}$ and (**g**–**l**) $F_{vdW}$.

### 4.2.1   $F_{EDL}$ for c–c, s–f, s–s, and cf–f models

The effects of systematic errors on the calculations and interpretations of $F_{EDL}$ for c–c, s–f, s–s, and cf–f models are shown in Fig. 12 and Fig. 13.

The main results are already summarized above. More specifically, for example, for c–c, s–f, and s–s models:

(1) $\Delta D$ of a few nm can drastically decrease $F_{EDL}$ (Fig. 12a) and the fitted $\psi/\sigma$ (Fig. 12b–c). At $|\psi_0| = 100$ mV and $c = 10^{-2}$ M, the ratios of $\psi'/\psi_0$ can be < 50% (at $\Delta D > 3.2$ nm) or even < 10% (at $\Delta D > 13$ nm) (Fig. 12b); at $|\sigma_0| = 0.05$ C m$^{-2}$ and $c = 10^{-2}$ M, the ratios of $\sigma'/\sigma_0$ can be < 50% (at $\Delta D > 1.5$ nm) or even < 10% (at $\Delta D > 8.5$ nm) (Fig. 12c) ($\psi'/\sigma'$ is the misinterpreted $\psi/\sigma$ and $\psi_0/\sigma_0$ the actual $\psi/\sigma$). And as $|\psi_0/\sigma_0|$, $\Delta D$, and $c$ increase, the ratios of $\psi'/\psi_0$ or $\sigma'/\sigma_0$ can become even smaller.

(2) The small relative errors of $R$ can drastically change $F_{EDL}$ (Fig. 12d) and the fitted $\psi/\sigma$ (Fig. 12e–f). At $|\sigma_0| = 0.05$ C m$^{-2}$ and $c = 10^{-2}$ M, a 17% positive error of $R$ ($E_R - 1 = 17\%$) can lead to $\sigma'/\sigma_0 < 50\%$ (Fig. 12f). And as $|\psi_0/\sigma_0|$ increases or $c$ decreases, the ratios of $\psi'/\psi_0$ or $\sigma'/\sigma_0$ can become even smaller as $E_R$ increases ($E_R - 1 > 0$). At $|\psi_0| = 150$ mV and any $c$, a 16% negative error of $R$ ($E_R - 1 = -16\%$) can lead to $\psi'/\psi_0 > 150\%$ and a 20% negative error of $R$ ($E_R - 1 = -20\%$) can even lead $\psi'$ to exceed the allowable range of calculation at any $c$ (Fig. 12e). At $|\sigma_0| = 0.05$ C m$^{-2}$ and $c = 10^{-3}$ M, a 5% negative error of $R$ ($E_R - 1 = -5\%$) can lead to $\sigma'/\sigma_0 > 150\%$ and a 14% negative error of $R$ ($E_R - 1 = -14\%$) can even lead $\sigma'$ to exceed the allowable range of calculation (Fig. 12f). And as $|\psi_0/\sigma_0|$ increases or $c$ decrease, the ratios of $\psi'/\psi_0$ or $\sigma'/\sigma_0$ can become even larger as $E_R$ decreases ($E_R - 1 < 0$). The descriptions of the results for $k$ (Fig. 12g–i) are similar to those of $R$ and are therefore omitted.

For cf–f model, the results are very similar. In fact, Fig. 13b–c and Fig. 13k–l are exactly the same as Fig. 12b–c and Fig. 12h–i, respectively. Fig. 13e–f and Fig. 13h–i are very similar to Fig. 12e–f. Therefore, the detailed descriptions are omitted.

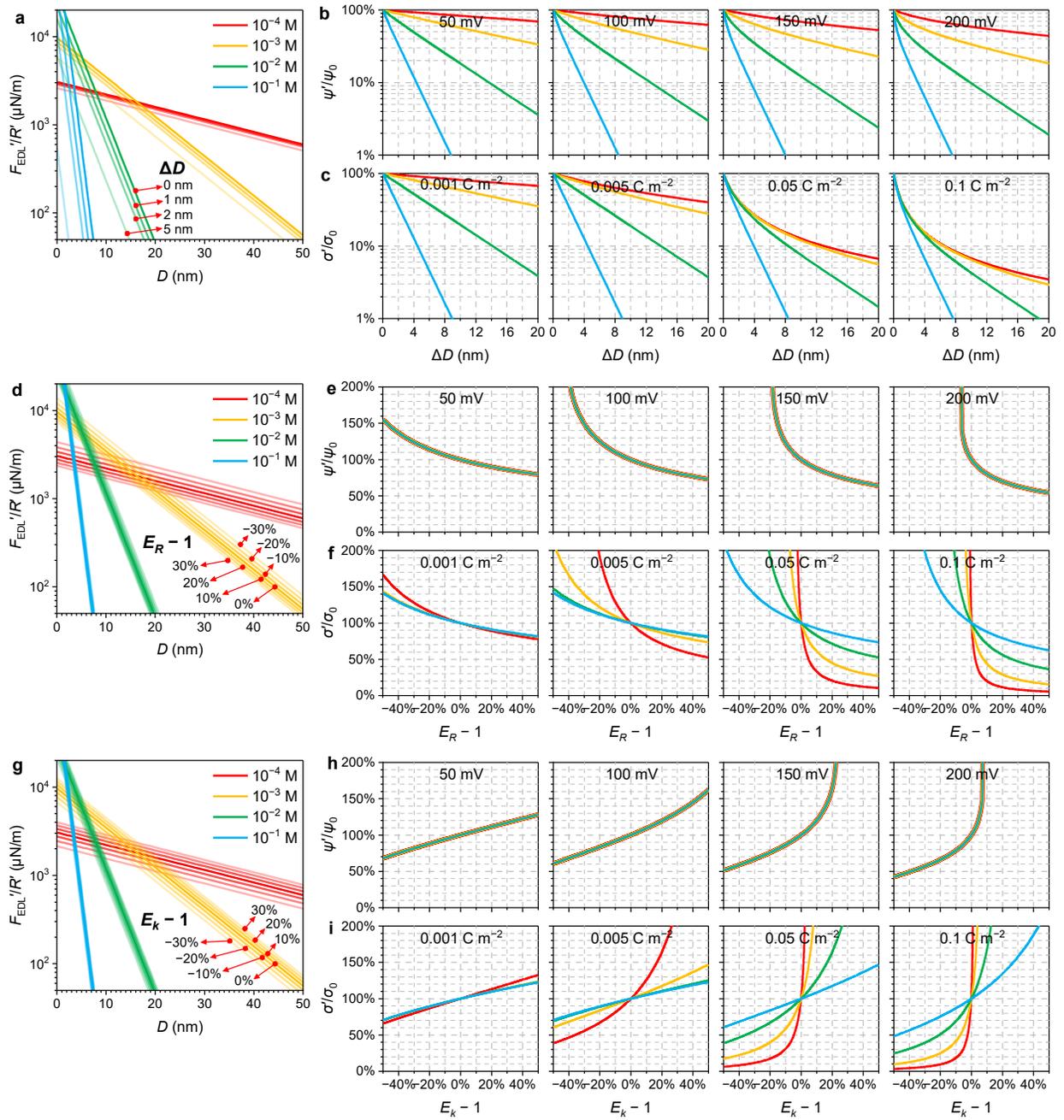

Fig. 12 Effects of systematic errors on the calculations and interpretations of $F_{EDL}$ for c–c, s–f, and s–s models.

(**a**–**c**) Effects of $\Delta D$ (set $E_R = E_k = 1$). (**a**) The theoretical $F_{EDL}'/R'$ curves with considering different $\Delta D$ (set $t = 1$). The ratios of (**b**) $\psi'/\psi_0$ and (**c**) $\sigma'/\sigma_0$ varying with $\Delta D$ at different $\psi_0/\sigma_0$ and $c$.

(**d**–**f**) Effects of $E_R - 1$ (set $\Delta D = 0$ and $E_k = 1$). (**d**) The theoretical $F_{EDL}'/R'$ curves with considering different $E_R - 1$ (set $t = 1$). The ratios of (**e**) $\psi'/\psi_0$ and (**f**) $\sigma'/\sigma_0$ varying with $E_R - 1$ at different $\psi_0/\sigma_0$ and $c$.

(**g**–**i**) Effects of $E_k - 1$ (set $\Delta D = 0$ and $E_R = 1$). (**d**) The theoretical $F_{EDL}'/R'$ curves with considering different $E_k - 1$ (set $t = 1$). The ratios of (**h**) $\psi'/\psi_0$ and (**i**) $\sigma'/\sigma_0$ varying with $E_k - 1$ at different $\psi_0/\sigma_0$ and $c$.

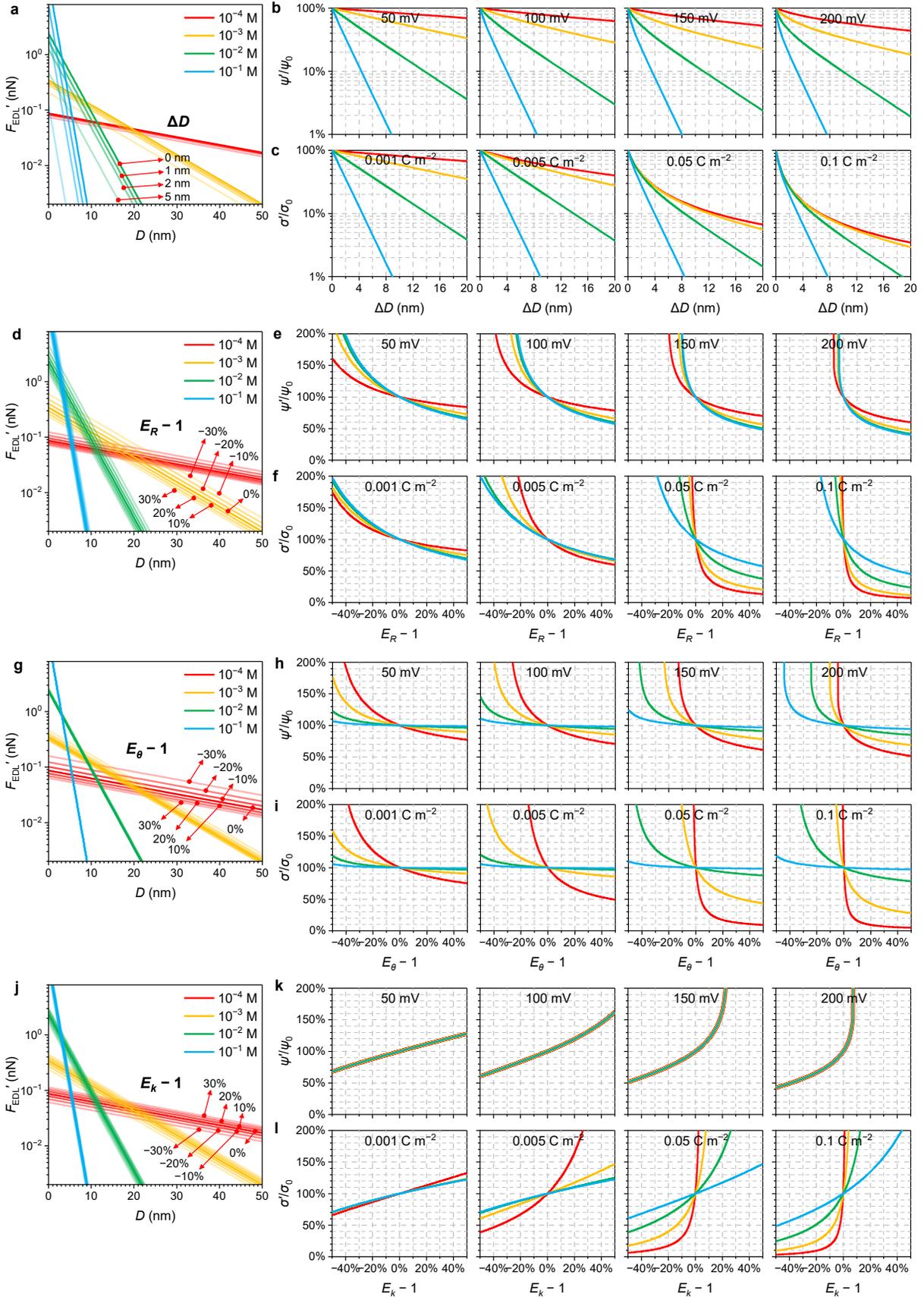

Fig. 13 Effects of systematic errors on the calculations and interpretations of $F_{EDL}$ for cf–f model (set $R_0$ =20 nm and $\theta_0 = 30°$).

(**a–c**) Effects of $\Delta D$ (set $E_R = E_\theta = E_k = 1$). (**a**) The theoretical $F_{EDL}'$ curves with considering different $\Delta D$ (set $t = 1$). The ratios of (**b**) $\psi'/\psi_0$ and (**c**) $\sigma'/\sigma_0$ varying with $\Delta D$ at different $\psi_0/\sigma_0$ and $c$.

(**d–f**) Effects of $E_R - 1$ (set $\Delta D = 0$ and $E_\theta = E_k = 1$). (**d**) The theoretical $F_{EDL}'$ curves with considering different $E_R - 1$ (set $t = 1$). The ratios of (**e**) $\psi'/\psi_0$ and (**f**) $\sigma'/\sigma_0$ varying with $E_R - 1$ at different $\psi_0/\sigma_0$ and $c$.

(**g–i**) Effects of $E_\theta - 1$ (set $\Delta D = 0$ and $E_R = E_k = 1$). (**d**) The theoretical $F_{EDL}'$ curves with considering different $E_\theta - 1$ (set $t = 1$). The ratios of (**h**) $\psi'/\psi_0$ and (**i**) $\sigma'/\sigma_0$ varying with $E_\theta - 1$ at different $\psi_0/\sigma_0$ and $c$.

(**j–l**) Effects of $E_k - 1$ (set $\Delta D = 0$ and $E_R = E_\theta = 1$). (**j**) The theoretical $F_{EDL}'$ curves with considering different $E_k - 1$ (set $t = 1$). The ratios of (**k**) $\psi'/\psi_0$ and (**l**) $\sigma'/\sigma_0$ varying with $E_k - 1$ at different $\psi_0/\sigma_0$ and $c$.

### 4.2.2  $F_{vdW}$ for c–c, s–f, s–s, and cf–f models

The effects of systematic errors on the calculations and interpretations of $F_{vdW}$ for c–c, s–f, and s–s models are shown in Fig. 14.

The main results are already summarized above. More specifically, for example, at $\Delta D = 5$ nm, the ratios of $H'/H_0$ ($H'$ is the misinterpreted $H$ and $H_0$ the actual $H$) can be < 50% (at $D < 12$ nm) or even < 10% (at $D < 2$ nm) (Fig. 14c).

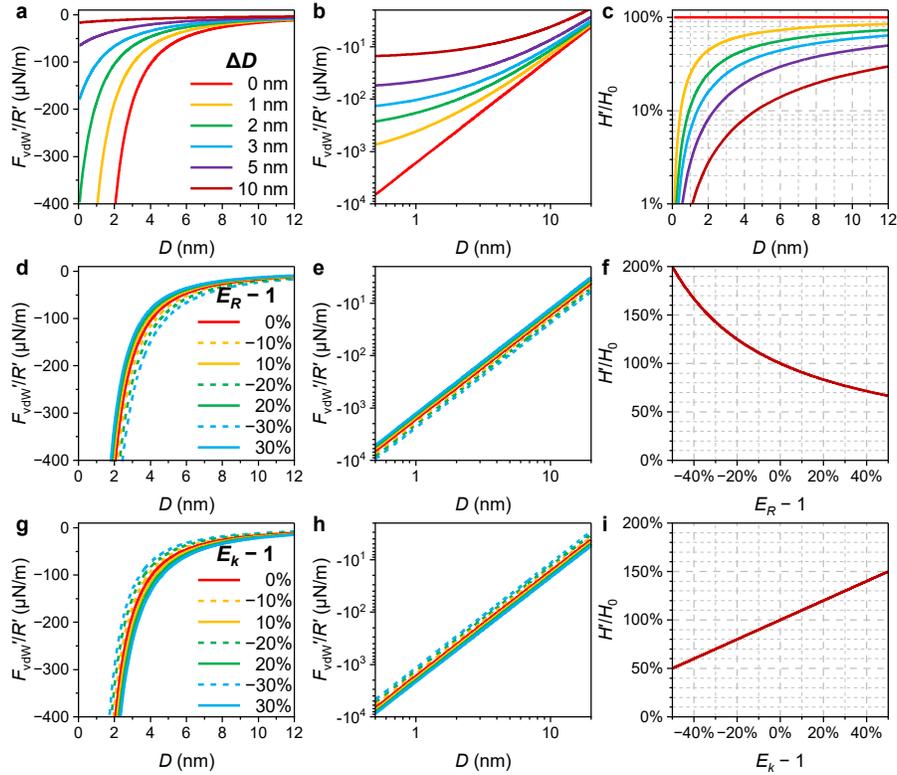

Fig. 14 Effects of systematic errors on the calculations and interpretations of $F_{vdW}$ for c–c, s–f, and s–s models.

(**a–c**) Effects of $\Delta D$ (set $E_R = E_k = 1$). The theoretical $F_{vdW}'/R'$ curves with considering different $\Delta D$ showed in (**a**) normal and (**b**) logarithmic $xy$-axis (set $H = 10^{20}$ J). (**c**) The ratios of $H'/H_0$ varying with $D$ at different $\Delta D$.

(**d–f**) Effects of $E_R - 1$ (set $\Delta D = 0$ and $E_k = 1$). The theoretical $F_{vdW}'/R'$ curves with considering different $E_R - 1$ showed in (**d**) normal and (**e**) logarithmic $xy$-axis (set $H = 10^{20}$ J). (**f**) The ratios of $H'/H_0$ varying with $D$ at different $E_R - 1$.

(**g–i**) Effects of $E_k - 1$ (set $\Delta D = 0$ and $E_R = 1$). The theoretical $F_{vdW}'/R'$ curves with considering different $E_k - 1$ showed in (**g**) normal and (**h**) logarithmic $xy$-axis (set $H = 10^{20}$ J). (**i**) The ratios of $H'/H_0$ varying with $D$ at different $E_k - 1$.

The effects of systematic errors on the calculations and interpretations of $F_{vdW}$ for cf–f model are shown in Fig. 15. The main results are similar to but somewhat complicated than that of $F_{vdW}$ for c–c, s–f, and s–s models. The details are as follows.

As pointed out in section 3.2, as $D \ll R$, $-F_{vdW}$ is approximately linearly related to $D$ in logarithmic $xy$-axis (dashed line in Fig. 15b). Fig. 15b also shows that the curvature of the $-F_{vdW}$ curve increases as $\Delta D$ increases (in logarithmic $xy$-axis).

The effect of $\Delta D$ (Fig. 15a–c) is significant (very similar to Fig. 14 a–c), while the effect of $E_R$, $E_\theta$, or $E_k$ is generally not significant than that of $\Delta D$. Except for $E_R$, which can also cause large errors of $H$ with positive errors of $R$ (Fig. 15f).

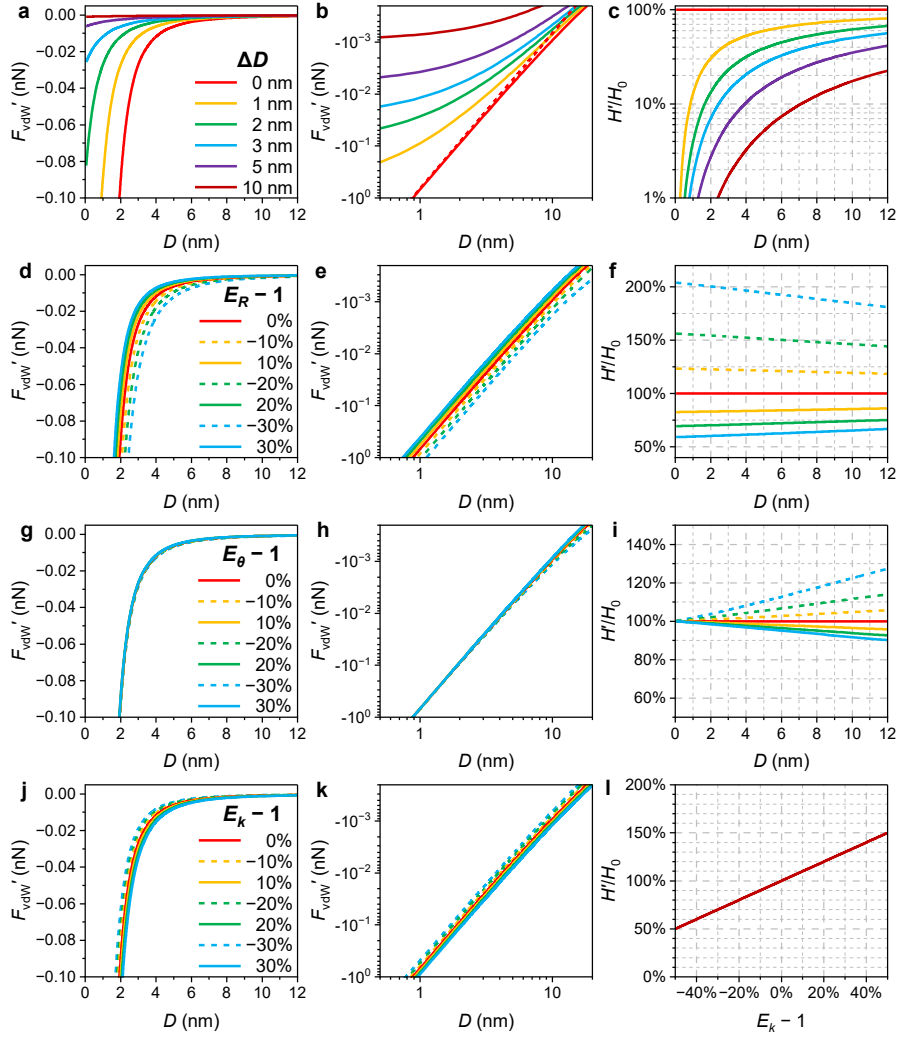

Fig. 15 Effects of systematic errors on the calculations and interpretations of $F_{vdW}$ for cf–f model (set $R_0$ =20 nm and $\theta_0$= 30°).

(**a**–**c**) Effects of $\Delta D$ (set $E_R = E_\theta = E_k = 1$). The theoretical $F_{vdW}'$ curves with considering different $\Delta D$ showed in (**a**) normal and (**b**) logarithmic *xy*-axis (set $H = 10^{20}$ J). (**c**) The ratios of $H'/H_0$ varying with $D$ at different $\Delta D$.

(**d**–**f**) Effects of $E_R - 1$ (set $\Delta D = 0$ and $E_\theta = E_k = 1$). The theoretical $F_{vdW}'$ curves with considering different $E_R - 1$ showed in (**d**) normal and (**e**) logarithmic *xy*-axis (set $H = 10^{20}$ J). (**f**) The ratios of $H'/H_0$ varying with $D$ at different $E_R - 1$.

(**g**–**i**) Effects of $E_\theta - 1$ (set $\Delta D = 0$ and $E_R = E_k = 1$). The theoretical $F_{vdW}'$ curves with considering different $E_\theta - 1$ showed in (**g**) normal and (**h**) logarithmic *xy*-axis (set $H = 10^{20}$ J). (**i**) The ratios of $H'/H_0$ varying with $D$ at different $E_\theta - 1$.

(**j**–**l**) Effects of $E_k - 1$ (set $\Delta D = 0$ and $E_R = E_\theta = 1$). The theoretical $F_{vdW}'$ curves with considering different $E_k - 1$ showed in (**j**) normal and (**k**) logarithmic *xy*-axis (set $H = 10^{20}$ J). (**l**) The ratios of $H'/H_0$ varying with $D$ at different $E_k - 1$.

### 4.3  Why are systematic errors commonly neglected?

So far, we have clearly illustrated that the effects of systematic errors are significant. In fact, the existence of systematic errors is easy to understand. We are aware that systematic errors always exist, because there are always deviations between the real experimental situations from the ideal theoretical situations (actual used parameters from ideal calculation models). However, the systematic errors are commonly neglected in the literature. Here, we try to explain why systematic errors are commonly neglected.

(1) The implicit characteristics of the effects of systematic errors on DLVO forces are commonly neglected.

The linear characteristic of $F_{EDL}$ (in logarithmic $y$-axis) can always be maintained with different $\psi/\sigma$ and different systematic errors (Fig. 11a–c). This means that, even with the presence of varying degrees of systematic errors, there may still be a corresponding $\psi/\sigma$ that makes a theoretical $F_{EDL}$ curve agrees well with an experimental $F_{EDL}$ curve (Fig. 11d–f). Such fitting results seem to be consistent with the DLVO theory, but the fitted $\psi/\sigma$ is far from the actual situation.

Similar for $F_{vdW}$, the linear characteristic of $-F_{vdW}$ (in logarithmic $xy$-axis) can always be maintained with different $H$ and different systematic errors ($E_R$ and $E_k$) (Fig. 11g and i). Besides, the characteristic of the original theoretical $-F_{vdW}$ curve changing from a straight line to a curve under the effect of $\Delta D$ can only be evident when using logarithmic $xy$-axis (Fig. 11h). In fact, with careful observation, this characteristic can be clearly found in the literature but has been entirely neglected [38,79].

Therefore, these implicit characteristics of the effects of systematic errors can lead to arbitrariness in the fitting results but are commonly neglected.

(2) The strong amplifying effects on systematic errors are commonly neglected.

As illustrated in section 4.2, the interpretation results (e.g., the fitted $\psi/\sigma/H$) have strong amplifying effects on systematic errors, which means that very small systematic errors can cause great deviations the interpretation results. However, without calculating and plotting, we may hardly believe that the effects of systematic errors can be such significant. Taking surface roughness as an example, many researchers may automatically assume that the effect of nanoscale roughness can be neglected because $F_{EDL}$ is a long-range force. This is clearly not true as demonstrated in the calculation results.

(3) The understandings of effects of systematic errors are usually inappropriate and incomplete.

Actually, as mentioned in section 4.1, some researchers are aware of the effects and importance of systematic errors but do not provide appropriate and complete assessments and corresponding effective solutions.

Taking surface roughness as an example, the effects of surface roughness have been noticed by some researchers. However, most of the researchers applied $\sigma_{RMS}$ rather than $H_{p-v}$ to evaluate and correct the effects of surface roughness [20,35,41]. This is inappropriate as highlighted in section 4.1.2 and proved in section 6. In fact, the effects of surface roughness can be quite complicated, but overly detailed consideration can be laborious and can obscure the key features [40,80]. Therefore, most researchers simply neglected systematic errors. Besides, given the diverse sources of systematic errors (see section 4.1) and the absence of appropriate correction methods, these studies are quite incomplete.

By contrast, in this section, we have identified and assessed the effects of systematic errors in a concise and simple manner, which can well reveal the characteristics of effects of systematic errors. Based on the revealed characteristics, we have further presented feasible solutions for correcting systematic errors in section 6, which also proves that the consideration and the revealed characteristics of systematic errors are reasonable and valid.

It is worth mentioning that, based on the effects of systematic errors, many other related phenomena can be identified and reasonably explained, and some related applications can also be developed. For example, owing to the effects of surface roughness, the spectacular oscillatory mode of the hydration force between atomically smooth surfaces will disappear with just a tiny roughness of the surfaces [81]. As also indicated by Valmacco et al. [38], controlling the surface roughness would open the possibility of tuning interaction forces precisely and have major implications in various research area, such as controlling the stability and self-assembly of colloidal suspensions [60,82,83] and fabricating micro-electromechanical systems (MEMS) based devices [84].

## 5  Revisiting the measurements and interpretations of DLVO forces

As mentioned in section 1.2, although numerous studies have been conducted in measurements and interpretations of DLVO forces, pervasive anomalous results can be identified throughout the literature.

Based on the above arguments about EDL theory and systematic errors (sections 2 and 4), we can reasonably–explain the pervasive anomalous results in the literature (section 5.1). We can further reasonably speculate that, the pervasive anomalous results are existing but not noticed and questioned in the literature owing to the two important aspects: (1) the pervasive unreasonable understandings of EDL theory and (2) the commonly neglected systematic errors (section 5.2). The details are as follows.

## 5.1 Reasonable explanations for the pervasive anomalous results

As explained in section 2.6, for the commonly used materials, the $\psi/\sigma$ range making $|F_{EDL}|$ tend to be maximum and constant (around $|\psi| \geq 150$ mV or $|\sigma| \geq 0.05$ C m$^{-2}$) can be easily satisfied at appropriate $c$ and pH.

However, as mentioned in section 1.2, the fitted $\psi/\sigma$ is normally extremely small and varies greatly throughout the literature. These anomalous results are pervasive in the related research of surface force measurements and are obviously contrary to the basic understandings of EDL theory as clarified in section 2. In fact, as predicted in section 4, the existence of $\Delta D$, positive errors of $R$ and $\theta$, and negative errors of $k$ can cause the fitted $\psi/\sigma$ to be extremely small (smaller than 50% or even 10% of the actual values). While the varying systematic errors in different experiments can cause the fitted $\psi/\sigma$ to vary greatly.

As mentioned in section 1.2, the fitted $\psi/\sigma$ can exceed the allowable range of calculation in the literature. In fact, as predicted in section 4, the existence of negative errors of $R$ and positive errors of $k$ can cause the fitted $\psi/\sigma$ to exceed the allowable range of calculation.

It is also worth noting that these phenomena are only evident when over-approximated $F_{EDL}$ formulas are used, because the $F_{EDL}$ calculated by the normal $F_{EDL}$ formulas (like LSA and exact calculation methods) has the "maximum and constant" limit (section 2.6), beyond which the fitted $\psi/\sigma$ can never be obtained. In this case, one would have to discard such anomalous data and repeat the experiments to obtain "reasonable" results.

As mentioned in section 1.2, the measured $F_{vdW}$ and the fitted $H$ vary greatly in the literature. In fact, as predicted in section 4, the existence of $\Delta D$, positive errors of $R$, and negative errors of $k$ can significantly decrease the measured $F_{vdW}$ and the fitted $H$. In particular, the existence of excessive surface roughness (characterized by large $\Delta D$) can even cause the measured $F_{vdW}$ to be undetectable (also showed in ref. [38]). Besides, because the systematic errors vary in different experiments, the measured DLVO forces ($F_{EDL}$ and $F_{vdW}$) and the fitted $\psi/\sigma/H$ under similar experimental conditions can vary greatly.

If we carefully observe the experimental $F_{vdW}$–$D$ curves, we can also find that they are usually not linearly related to $D$ in logarithmic $xy$-axis [38,79]. This is obviously inconsistent with the important "linear" characteristic of $F_{vdW}$ (Eq. (8) and as depicted in section 3.2) and has been commonly neglected by researchers. Actually, as predicted in section 4, the existence of $\Delta D$ can change the original theoretical $-F_{vdW}$ curve from a straight line to a curve (Fig. 11h) (in logarithmic $xy$-axis).

## 5.2 Why are the pervasive anomalous results not noticed and questioned?

Based on section 5.1, we can reasonably infer that the above anomalous results are caused by the neglected systematic errors. However, although the pervasive anomalous results can almost be found everywhere, most researchers gradually accept the anomalous results, without noticing and questioning. Therefore, we try to explain the underlying reasons why the pervasive anomalous results are not noticed and questioned in measurements and interpretations of DLVO forces (especially $F_{EDL}$).

It's extremely important for us to realize that there are two prerequisites for theory and experiments to coincide: (1) the theoretical assumptions are reasonable; and (2) the imperfections of experimental systems are considered.

Obviously, the systematic errors in experimental systems can directly cause the anomalous results as summarized in section 5.1. The anomalous results can be used as strong evidence for the discrepancies between experiment and theory and the existence of systematic errors. However, few researchers have mentioned and evaluated the systematic errors. This is largely because, as illustrated in section 4.3, the systematic errors are commonly neglected.

Besides, the anomalous results also reflect that the theoretical assumptions of EDL theory clarified in section 2 are misunderstood. However, few researchers have raised questions about the unreasonable understandings. This is largely because the pervasive unreasonable understandings of EDL theory (summarized in section 2.7) have always existed and been reinforced by the pervasive anomalous results (summarized in section 5.1).

Therefore, we can conclude that it is the commonly neglected systematic errors (section 4) that have directly led to deviations between experiments and theory and have reinforced the unreasonable understandings of EDL theory (section 2); while due to the pervasive unreasonable understandings of EDL theory, researchers have gradually accepted the anomalous results in measurements and interpretations of DLVO forces. Once we reasonably consider the EDL theory and the systematic errors, the experiments will naturally coincide with the theory.

## 6  Correction of systematic errors

In order to obtain reasonable results in measurements and interpretations of DLVO forces, it is necessary to apply the EDL theory appropriately and deduct the systematic errors. Based on the characteristics of $F_{EDL}$ (section 2.6), $|F_{EDL}|$ has the important characteristic that it tends to be maximum and constant and independent of $\psi/\sigma$ when $\psi/\sigma$ is large enough at appropriate $c$ and pH. While according to the characteristics of systematic errors (section 4), being systematic errors implies that their magnitude and effects are exactly the same in the same batch of experiments. Therefore, the systematic error parameters can possibly be calculated and deducted by comparing the difference between the calculated theoretical $F_{EDL}$ curves and the experimental $F_{EDL}$ curves.

In section 6, based on the characteristics of EDL theory (section 2) and systematic errors (section 4), we further construct theoretical (section 6.1) and experimental methods (section 6.2) and provide experimental verifications (section 6.3) for correcting systematic errors in measurements and interpretations of DLVO forces.

The results can well prove that: (1) The systematic errors are universal and do have significant effects on the fitted results, and therefore must be corrected prior to the experiments. By correcting systematic errors, reasonable understandings of EDL theory can be obtained in experiments. (2) The clarifications on the EDL theory in section 2 and the identification and effects analyses of systematic errors in section 4 are reasonable and valid.

The details are as follows.

### 6.1  Theoretical calculation processes for correction of systematic errors

In this section, we construct theoretical methods for correcting systematic errors in measurements and interpretations of DLVO forces. Theoretically, by constructing simple theoretical calculation processes, we only need to measure a few force curves at different $c$ and ensure that the $\psi/\sigma$ is large enough, all the systematic error parameters for the three surface force measurement techniques (section 4.1) can be obtained, as illustrated in Fig. 16. The detailed theoretical methods are as follows.

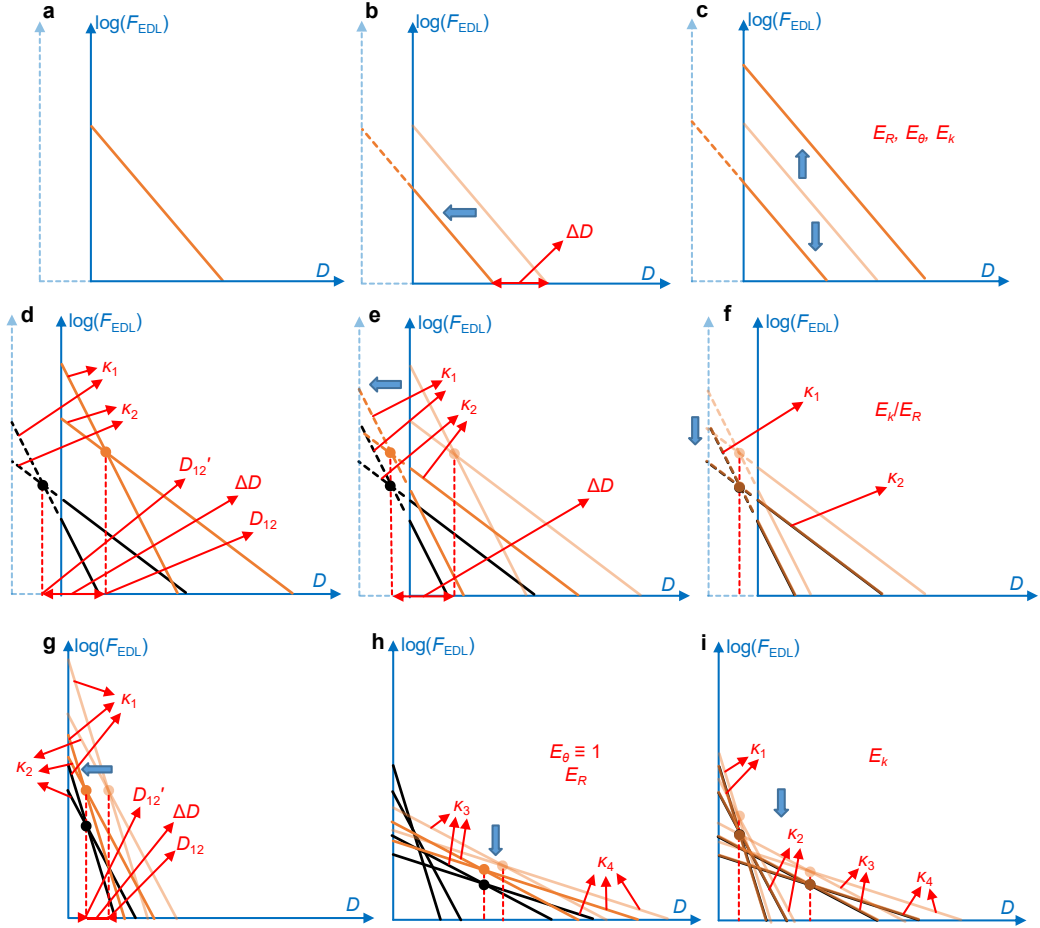

Fig. 16 Schematic of the theoretical methods for correcting systematic errors (in logarithmic $y$-axis). (**a–c**) Changes of a theoretical $F_{EDL}$ curve after considering systematic errors. (**d–f**) The process of calculating the systematic error parameters for SFA and AFM microsphere probe technique. (**g–i**) The process of calculating the systematic error parameters for AFM nanoscale tip technique.

### 6.1.1 SFA and AFM microsphere probe technique

As depicted in section 4.1, three systematic error parameters, $\Delta D$, $E_R$, and $E_k$, should be used to quantify the effects of systematic errors for c–c, s–f, and s–s models (Fig. 1d–f).

When the $\psi/\sigma$ is large, we have $t(t_1 t_2) \rightarrow 1$. Based on Eq. (6), the original theoretical $F_{EDL}$ curves without considering systematic error parameters can be expressed as (represented by Fig. 16a)

$$F_{\text{EDL(The)}} = A\kappa R' e^{-\kappa D} \Leftrightarrow \ln F_{\text{EDL(The)}} = -\kappa D + \ln(A\kappa R') \tag{11}$$

While the final new theoretical $F_{\text{EDL}}$ curves with considering systematic error parameters $\Delta D$, $E_R$, and $E_k$, which are consistent with the experimental $F_{\text{EDL}}$ ($F_{\text{EDL(Exp)}}$) curves, should be expressed as

$$F_{\text{EDL(The)}}' = E_k A\kappa (R'/E_R) e^{-\kappa(D+\Delta D)} \Leftrightarrow \ln F_{\text{EDL(The)}}' = -\kappa(D+\Delta D) + \ln\left[E_k A\kappa(R'/E_R)\right] \tag{12}$$

And Eq. (12) is equivalent to move the original theoretical $F_{\text{EDL}}$ curves (expressed by Eq. (11)) along the $x$-axis to the left by $\Delta D$ (Fig. 16b) and along the $y$-axis up/down (Fig. 16c) in logarithmic $y$-axis, respectively.

By comparing the difference between the $F_{\text{EDL(The)}}$ curves (expressed by Eq. (11)) and the $F_{\text{EDL(Exp)}}$ curves (represented by Eq. (12)), the three systematic error parameters, $\Delta D$, $E_R$, and $E_k$, can be directly calculated. However, comparing Fig. 16b and Fig. 16c, it can be seen that actually we cannot distinguish all systematic error parameters by a single force curve.

While two $F_{\text{EDL(Exp)}}$ curves (represented by Eq. (12)), with different $\kappa$ of $\kappa_i$ and $\kappa_j$ (or $c$ of $c_i$ and $c_j$), and the corresponding two $F_{\text{EDL(The)}}$ curves without considering systematic error parameters (expressed by Eq. (11)) have their respective intersection points in logarithmic $y$-axis. Therefore, we have

$$\begin{cases} \ln F_{\text{EDL(The)}}'^{i} = -\kappa_i(D+\Delta D) + \ln\left[E_k A\kappa_i(R'/E_R)\right] \\ \ln F_{\text{EDL(The)}}'^{j} = -\kappa_j(D+\Delta D) + \ln\left[E_k A\kappa_j(R'/E_R)\right] \end{cases} \tag{13}$$

$$\begin{cases} \ln F_{\text{EDL(The)}}^{i} = -\kappa_i D + \ln(A\kappa_i R') \\ \ln F_{\text{EDL(The)}}^{j} = -\kappa_j D + \ln(A\kappa_j R') \end{cases} \tag{14}$$

Solving Eqs. (13) and (14), respectively, the abscissas of the two intersection points can be obtained as

$$D_{ij}' = -\Delta D + \frac{1}{\kappa_i - \kappa_j}\ln\frac{\kappa_i}{\kappa_j} \tag{15}$$

$$D_{ij} = \frac{1}{\kappa_i - \kappa_j}\ln\frac{\kappa_i}{\kappa_j} \tag{16}$$

(1) Calculation of $\Delta D$.

As shown in Fig. 16d, with two known $\kappa$ of $\kappa_1$ and $\kappa_2$ (or $c$ of $c_1$ and $c_2$), we can obtain $\Delta D$, which equals the horizontal distance of the two intersection points of the two $F_{\text{EDL(Exp)}}$ curves (black lines, represented by Eq. (12)) and the corresponding two $F_{EDL(The)}$ curves without considering systematic error parameters (orange lines, expressed by Eq. (11)), by comparing Eqs. (15) and (16)

$$\Delta D = D_{12} - D_{12}' \tag{17}$$

Introducing the obtained $\Delta D$ into Eq. (11), we get the new theoretical $F_{EDL}$ curves with considering only $\Delta D$

$$\ln F_{EDL(The)}'' = -\kappa(D + \Delta D) + \ln(A\kappa R') \tag{18}$$

And Eq. (18) (orange lines) is equivalent to move Eq. (11) (light orange lines) along the x-axis to the left by $\Delta D$, as shown in Fig. 16e.

(2) Calculation of $E_R$ and $E_k$.

At this moment, it is found in Fig. 16e that the $F_{EDL(The)}''$ (Eq. (18), orange lines) and $F_{EDL(Exp)}$ (represented by Eq. (12), black lines) curves are still offset along the y-axis, indicating the existence of $E_R$ and $E_k$. Further comparing Eq. (18) and Eq. (12), the whole $E_k/E_R$ can be obtained ($E_R$ and $E_k$ cannot be distinguished separately, and there is no need to distinguish them).

Eq. (12) can be obtained by further introducing $E_k/E_R$ into Eq. (18), which means the final $F_{EDL(The)}'$ curves are in perfect agreement with the $F_{EDL(Exp)}$ curves (an overlay of orange and black lines), as shown in Fig. 16f.

### 6.1.2 AFM nanoscale tip technique

Because of the more complex $F_{EDL}$ expressions and additional systematic error parameters, the calculation process is much more complicated than that in section 6.1.1.

As depicted in section 4.1, four systematic error parameters, $\Delta D$, $E_R$, $E_\theta$, and $E_k$, should be used to quantify the effects of systematic errors for cf–f model (Fig. 1g).

When the $\psi/\sigma$ is large, we have $t(t_1t_2) \to 1$. Based on Eq. (9), the original theoretical $F_{EDL}$ curves without considering systematic error parameters can be expressed as (represented by Fig. 16a)

$$F_{e(The)} = A\left(\frac{1}{2}\kappa^2 R'^2 + \tan^2\theta' + \kappa R' \tan\theta'\right)e^{-\kappa D}$$
$$\Leftrightarrow \ln F_{e(The)} = -\kappa D + \ln\left[A\left(\frac{1}{2}\kappa^2 R'^2 + \tan^2\theta' + \kappa R' \tan\theta'\right)\right] \tag{19}$$

While the final new theoretical $F_{EDL}$ curves with considering systematic error parameters $\Delta D$, $E_R$, $E_\theta$, and $E_k$, which are consistent with the experimental $F_{EDL}$ ($F_{EDL(Exp)}$) curves, should be expressed as

$$F_{EDL(The)}' = E_k A \left[\frac{1}{2}\kappa^2 \left(R'/E_R\right)^2 + \tan^2\left(\theta'/E_\theta\right) + \kappa\left(R'/E_R\right)\tan\left(\theta'/E_\theta\right)\right] e^{-\kappa(D+\Delta D)}$$

$$\Leftrightarrow \ln F_{EDL(The)}' = -\kappa(D+\Delta D) + \ln\left\{E_k A\left[\frac{1}{2}\kappa^2 \left(R'/E_R\right)^2 + \tan^2\left(\theta'/E_\theta\right) + \kappa\left(R'/E_R\right)\tan\left(\theta'/E_\theta\right)\right]\right\}$$

(20)

And Eq. (20) is equivalent to move the original theoretical $F_{EDL}$ curves (Eq. (19)) along the x-axis to the left by $\Delta D$ (Fig. 16b) and along the y-axis up/down (Fig. 16c) in logarithmic y-axis, respectively.

Similar to section 6.1.1, two $F_{EDL(Exp)}$ curves (represented by Eq. (20)), with different $\kappa$ of $\kappa_i$ and $\kappa_j$ (or $c$ of $c_i$ and $c_j$), and the corresponding two $F_{EDL(The)}$ curves without considering systematic error parameters (expressed by Eq. (19)) have their respective intersection points in logarithmic y-axis. Therefore, we have

$$\begin{cases} \ln F_{EDL(The)}'^{i} = -\kappa_i(D+\Delta D) + \ln\left\{E_k A\left[\frac{1}{2}\kappa_i^2\left(R'/E_R\right)^2 + \tan^2\left(\theta'/E_\theta\right) + \kappa_i\left(R'/E_R\right)\tan\left(\theta'/E_\theta\right)\right]\right\} \\ \ln F_{EDL(The)}'^{j} = -\kappa_j(D+\Delta D) + \ln\left\{E_k A\left[\frac{1}{2}\kappa_j^2\left(R'/E_R\right)^2 + \tan^2\left(\theta'/E_\theta\right) + \kappa_j\left(R'/E_R\right)\tan\left(\theta'/E_\theta\right)\right]\right\} \end{cases}$$

(21)

$$\begin{cases} \ln F_{EDL(The)}^{i} = -\kappa_i D + \ln\left[A\left(\frac{1}{2}\kappa_i^2 R'^2 + \tan^2\theta' + \kappa_i R'\tan\theta'\right)\right] \\ \ln F_{EDL(The)}^{j} = -\kappa_j D + \ln\left[A\left(\frac{1}{2}\kappa_j^2 R'^2 + \tan^2\theta' + \kappa_j R'\tan\theta'\right)\right] \end{cases}$$

(22)

Solving Eqs. (21) and (22), respectively, the abscissas of the two intersection points can be obtained as

$$D_{ij}' = -\Delta D + \frac{1}{\kappa_i - \kappa_j}\ln\frac{\kappa_i^2\left(R'/E_R\right)^2 + 2\tan^2\left(\theta'/E_\theta\right) + 2\kappa_i\left(R'/E_R\right)\tan\left(\theta'/E_\theta\right)}{\kappa_j^2\left(R'/E_R\right)^2 + 2\tan^2\left(\theta'/E_\theta\right) + 2\kappa_j\left(R'/E_R\right)\tan\left(\theta'/E_\theta\right)}$$ (23)

$$D_{ij} = \frac{1}{\kappa_i - \kappa_j}\ln\frac{\kappa_i^2 R'^2 + 2\tan^2\theta' + 2\kappa_i R'\tan\theta'}{\kappa_j^2 R'^2 + 2\tan^2\theta' + 2\kappa_j R'\tan\theta'}$$ (24)

(1) Calculation of $\Delta D$.

When $c$ is slightly high (while ensuring that the $\psi/\sigma$ is still large) so that $\kappa R \gg 1$, Eqs. (23) and (24) can be simplified as

$$D_{ij}' \approx -\Delta D + \frac{2}{\kappa_i - \kappa_j} \ln \frac{\kappa_i}{\kappa_j} \tag{25}$$

$$D_{ij} \approx \frac{2}{\kappa_i - \kappa_j} \ln \frac{\kappa_i}{\kappa_j} \tag{26}$$

As shown in Fig. 16g, with two known $\kappa$ of $\kappa_1$ and $\kappa_2$ (or $c$ of $c_1$ and $c_2$), we can obtain $\Delta D$, which equals the horizontal distance of the two intersection points of the two $F_{EDL(Exp)}$ curves (black lines, represented by Eq. (20)) and the corresponding two $F_{EDL(The)}$ curves without considering systematic error parameters (light orange lines, expressed by Eq. (19)), by comparing Eqs. (25) and (26)

$$\Delta D = D_{12} - D_{12}' \tag{27}$$

Introducing the obtained $\Delta D$ into Eq. (19), we get the new theoretical $F_{EDL}$ curves with considering only $\Delta D$

$$\ln F_{EDL(The)}'' = -\kappa(D + \Delta D) + \ln\left[A\left(\frac{1}{2}\kappa^2 R'^2 + \tan^2\theta' + \kappa R' \tan\theta'\right)\right] \tag{28}$$

And Eq. (28) (orange lines) is equivalent to move Eq. (19) (light orange lines) along the $x$-axis to the left by $\Delta D$, as shown in Fig. 16g.

(2) Calculation of $E_\theta$ and $E_R$.

When $c$ is low so that $\kappa R$ is close to or less than 1, Eq. (23) cannot be simplified as Eq. (25). Because the abscissa of the intersection point of two $F_{EDL(Exp)}$ curves (represented by $D_{ij}'$ in Eq. (23)) can be obtained in experiments, theoretically, the two unknowns in Eq. (23), $E_R$ and $E_\theta$, could be solved respectively by combining at least two sets of Eq. (23) with different $\kappa$ (or $c$).

However, it is found in the actual calculation that we cannot obtain a unique solution for both $E_R$ and $E_\theta$ through Eq. (23). The proof is as follows.

The solution of the two unknowns in Eq. (23), $E_R$ and $E_\theta$, is the intersection of the general solutions of at least two sets of Eq. (23) with different $\kappa$ (or $c$).

With any two different $\kappa$, $(\kappa_i, \kappa_j)$, assuming that the general solution to Eq. (23) is $(E_R^i, E_\theta^i)$ and one of the particular solutions is $(E_R^0, E_\theta^0)$, we have

$$D_{ij}' = -\Delta D + \frac{1}{\kappa_i - \kappa_j} \ln \frac{\kappa_i^2 \left( R'/E_{R^i} \right)^2 + 2\tan^2\left(\theta'/E_{\theta^i}\right) + 2\kappa_i \left( R'/E_{R^i} \right)\tan\left(\theta'/E_{\theta^i}\right)}{\kappa_j^2 \left( R'/E_{R^i} \right)^2 + 2\tan^2\left(\theta'/E_{\theta^i}\right) + 2\kappa_j \left( R'/E_{R^i} \right)\tan\left(\theta'/E_{\theta^i}\right)}$$

$$= -\Delta D + \frac{1}{\kappa_i - \kappa_j} \ln \frac{\kappa_i^2 \left( R'/E_{R^0} \right)^2 + 2\tan^2\left(\theta'/E_{\theta^0}\right) + 2\kappa_i \left( R'/E_{R^0} \right)\tan\left(\theta'/E_{\theta^0}\right)}{\kappa_j^2 \left( R'/E_{R^0} \right)^2 + 2\tan^2\left(\theta'/E_{\theta^0}\right) + 2\kappa_j \left( R'/E_{R^0} \right)\tan\left(\theta'/E_{\theta^0}\right)} \quad (29)$$

And Eq. (29) can be simplified as

$$\left( R'/E_{R^i} \right)\tan\left(\theta'/E_{\theta^0}\right) = \left( R'/E_{R^0} \right)\tan\left(\theta'/E_{\theta^i}\right) \quad (30)$$

Eq. (30) implies that, with any two different $\kappa$, $(\kappa_i, \kappa_j)$, the general solution to Eq. (23), $(E_R^i, E_\theta^i)$, is only related to the particular solution, $(E_R^0, E_\theta^0)$, and is independent of $(\kappa_i, \kappa_j)$. Since all sets of Eq. (23) with different $\kappa$ have at least one identical particular solution, for example, the real $E_R$ and $E_\theta$, they must also have the same general solution which satisfies Eq. (30). Therefore, we can actually never obtain a unique solution for both $E_R$ and $E_\theta$ through Eq. (23).

As a result, one of $\theta_0$ (= $\theta'/E_\theta$) and $R_0$ (= $R'/E_R$) must be obtained first, and another can be derived accordingly through Eq. (23).

The next question is whether one of the two, $\theta_0$ and $R_0$, can be obtained with relatively high accuracy by other means?

As explained in section 4.1 and our previous work [33], the two surface geometry parameters for AFM nanoscale tips, $R$ and $\theta$, are usually directly observed through SEM images and are highly inaccurate.

Specifically, for the equivalent tip end radius $R$, it is inappropriate to approximate the tip ends of commonly used pyramidal AFM tips as spherical, which means the $R$ observed through SEM images is also unreasonable [33,74]. After being smoothed into flat surfaces, the $R$ of AFM tips is also difficult to estimate accurately through SEM images due to the large error [33,85–87]. While for the equivalent circular cone half angle $\theta$ of pyramidal AFM tips, it actually can hardly be estimated through SEM images [33,74–76].

But in fact, the pyramidal AFM tips are commonly manufactured with a high degree of precision, which means the equivalent circular cone half angle $\theta$ can be calculated with relatively high accuracy by the tip specification provided from the manufacturers [33]. Taking the MLCT-BIO probes (Bruker) as an example (Fig. 17c4–c5 and SI section 3), the geometry of the tip is a square pyramid and the angle between its each face and the central axis (z-axis) is $35 \pm 2°$, with a relative uncertainty (relative error) of ±5.7%; while the calculated equivalent circular cone half angle $\theta$ is 39.8 ± 2.2°, with a relative uncertainty (relative error) of ±5.5%. According to the effects of systematic errors illustrated in section 4.2, the systematic error of $\theta$ can be reasonably ignored under certain conditions.

Therefore, we first assume that the calculated apparent $\theta$, $\theta'$, is approximately equal to the effective (actual) $\theta$, $\theta_0$, i.e., $E_\theta (= \theta'/\theta_0) \equiv 1$.

With the known $E_\theta$, $E_R (= R'/R_0)$ can be solved according to another set of Eq. (23) with two known $\kappa$ of $\kappa_3$ and $\kappa_4$ (or $c$ of $c_3$ and $c_4$) (the abscissa of the intersection of two $F_{EDL(Exp)}$ curves (represented by Eq. (20))):

$$D_{34}' = -\Delta D + \frac{1}{\kappa_3 - \kappa_4} \ln \frac{\kappa_3^2 \left(R'/E_R\right)^2 + 2\tan^2\left(\theta'/E_\theta\right) + 2\kappa_3 \left(R'/E_R\right)\tan\left(\theta'/E_\theta\right)}{\kappa_4^2 \left(R'/E_R\right)^2 + 2\tan^2\left(\theta'/E_\theta\right) + 2\kappa_4 \left(R'/E_R\right)\tan\left(\theta'/E_\theta\right)} \quad (31)$$

Further replacing $\theta'$ and $R'$ in Eq. (28) by $\theta'/E_\theta$ and $R'/E_R$, we get the new theoretical $F_{EDL}$ curves with considering $\Delta D$, $E_\theta$, and $E_R$

$$\ln F_{EDL(The)}''' = -\kappa(D+\Delta D) + \ln\left\{A\left[\frac{1}{2}\kappa^2 \left(R'/E_R\right)^2 + \tan^2\left(\theta'/E_\theta\right) + \kappa\left(R'/E_R\right)\tan\left(\theta'/E_\theta\right)\right]\right\} \quad (32)$$

And Eq. (32) (orange lines) is equivalent to move Eq. (28) (light orange lines) along the y-axis up/down by a certain value (the overall effect is to make the abscissas of the two intersections of the $F_{EDL(The)}'''$ (Eq. (32), orange lines) and the $F_{EDL(Exp)}$ (represented by Eq. (20), black lines) curves equal), as shown in Fig. 16h.

(3) Calculation of $E_k$.

At this moment, it is found in Fig. 16h that the $F_{EDL(The)}'''$ (Eq. (32), orange lines) and $F_{EDL(Exp)}$ (represented by Eq. (20), black lines) curves are still offset along the y-axis, indicating the existence of $E_k$. Further comparing Eq. (32) and Eq. (20), $E_k$ can be obtained.

Eq. (20) can be obtained by further introducing $E_k$ into Eq. (32), which means the final $F_{EDL(The)}'$ curves are in perfect agreement with the $F_{EDL(Exp)}$ curves (an overlay of orange and black lines), as shown in Fig. 16i.

## 6.2 Experimental principles for correction of systematic errors

Based on the constructed theoretical methods for correcting systematic errors in section 6.1, the corresponding experimental verifications for the three surface force measurement techniques (section 3.1) can be performed. The key to the successful application of the theoretical methods is to ensure that the $\psi/\sigma$ is large enough. Based on section 2.6, we chose the simplest experimental systems: common materials (mica, silica, and $Si_3N_4$) and solutions containing strongly hydrated counterions $Li^+$ with appropriate pH and $c$.

As pointed out in section 3.2, for $F_{EDL}$, we only need to fit the $D > \kappa^{-1}$ part of the $F_{DLVO}$–$D$ curves, and the $F_{vdW}$ generally have much shorter interaction range (a few nm) than $F_{EDL}$. Therefore, the measured force curves can generally be regarded as the same as $F_{EDL}$ curves.

The whole experiments are quite rigorous but still achievable. Special attentions should be paid to guarantee the reproducibility of the experimental data. The detailed experimental procedures for correcting systematic errors in the three surface force measurement techniques are depicted in SI section 3.

## 6.3 Experimental verifications

Based on the constructed theoretical methods for correcting systematic errors in section 6.1, the corresponding experimental verifications for the three surface force measurement techniques (section 3.1) can be performed according to section 6.2.

The results are demonstrated as follows.

### 6.3.1 Correcting systematic errors for SFA

For SFA, the most commonly used material is mica. The experimental data (data points, selected from ref. [15]) and correction results (black lines) of interaction forces between mica in $Li^+$ ($LiNO_3$) solutions in logarithmic $y$-axis are shown in Fig. 17a1 (in normal $xy$-axis in Fig. 18a). The $\psi/\sigma$ of mica can be guaranteed to be large enough at these conditions ($c$ and pH), as explained in section 2.6.

Fig. 17a1 shows that the theoretical $F_{EDL}$ curves after considering systematic error parameters at all $c$ are in good agreement with the experimental force curves, indicating the correctness of the correction method and results. The obtained systematic error parameters are $\Delta D = 0$ nm and $E_k/E_R = 0.75$.

Since mica has atomically smooth surface, it can be regarded as having no roughness. Therefore, $\Delta D$ should be 0 nm, which is exactly confirmed by the correction results. The whole $E_k/E_R = 0.75$, indicating that $E_k$ and $E_R$ together cause the experimental force curves to be 75% of the theoretical ones (without considering systematic error parameters). This can be explained by the measurement errors of $R$ and $k$ within the reported ranges of systematic errors in section 4.1.

As shown in Fig. 17a2–a3, the $\psi'$ and $\sigma'$ (misinterpreted $\psi$ and $\sigma$, fitted results for $\psi$ and $\sigma$ without considering systematic error parameters) for mica at all $c$ are very small. All $|\psi'|$ are < 150 mV (−133, −131, −123, and −104) and the corresponding $\sigma'$ are −0.016, −0.025, −0.071, −0.154 mC m$^{-2}$, respectively. In fact, we can only be certain that the values of $\psi_0$ and $\sigma_0$ (actual $\psi$ and $\sigma$, fitted results for $\psi$ and $\sigma$ with considering systematic error parameters) are large enough but cannot tell their exact values. Here, Fig. 17a2–a3 demonstrate the values of $\sigma_0 = \sigma_s = -0.34$ C m$^{-2}$ (see section 2.6) and the corresponding $\psi_0$ (−289, −265, −203, and −144 mV).

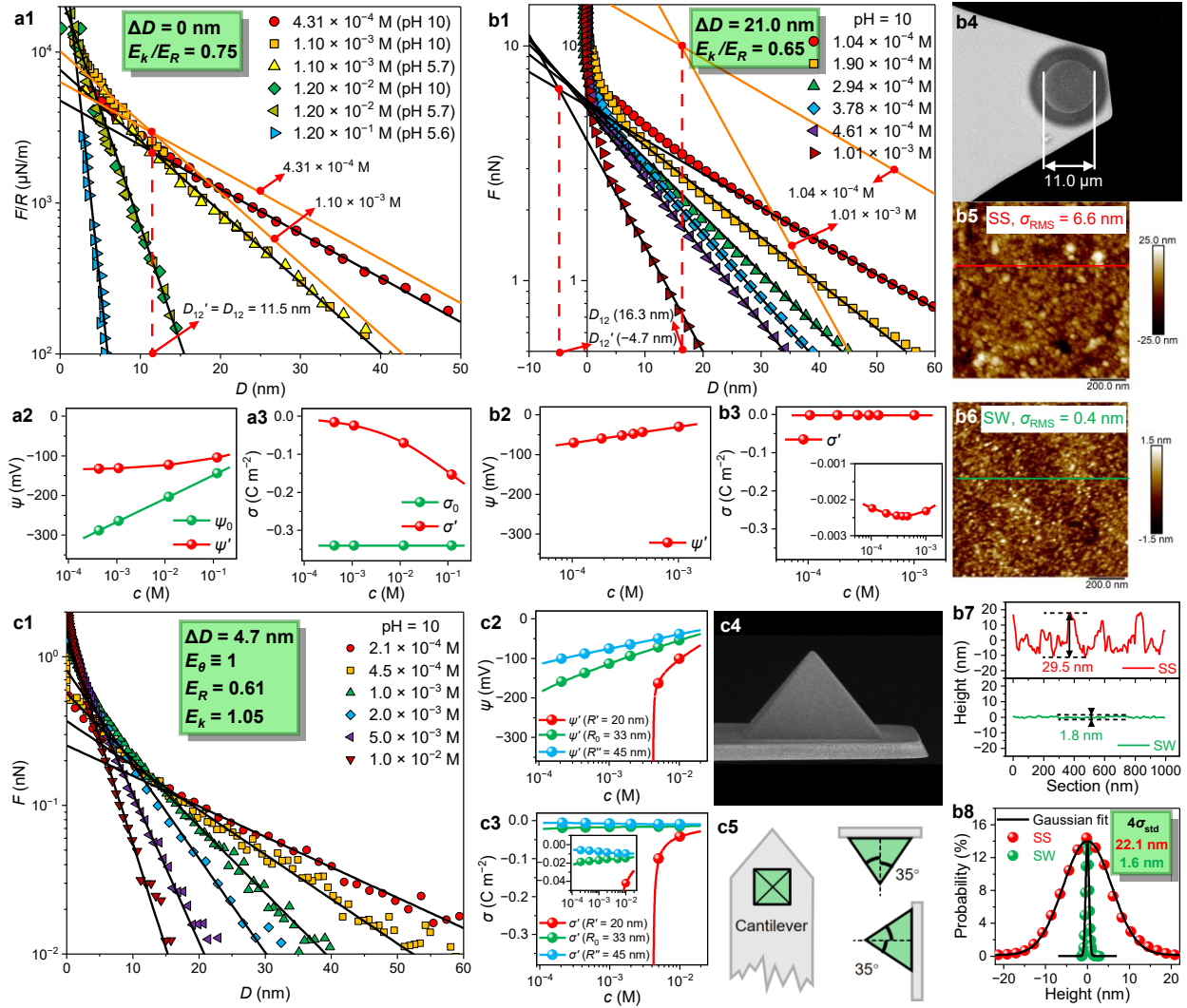

Fig. 17 Correcting systematic errors for the three surface force measurement techniques: (**a**) SFA, (**b**) AFM microsphere probe technique, and (**c**) AFM nanoscale tip technique.

(**a1**) Correction results of interaction forces between mica in $Li^+$ ($LiNO_3$) solutions measured by SFA. Data points (only 1/2 are shown for clarity) are selected from ref. [15]. The black and yellow lines are the theoretical $F_{EDL}$ curves with and without considering systematic error parameters, respectively. The $c$ in the legend are the best fitted results, and the $\kappa^{-1}$ corresponding to the four concentrations are 14.8, 9.26, 2.80, and 0.887 nm, respectively. Comparison of fitted results for (**a2**) $\psi$ and (**a3**) $\sigma$ with ($\psi_0$ and $\sigma_0$) and without ($\psi'$ and $\sigma'$) considering systematic error parameters.

(**b1**) Correction results of interaction forces between a silica microsphere (SS) and a silica wafer (SW) in $Li^+$ (LiCl) solutions at pH 10 measured by AFM microsphere probe technique. The data points (only 1/2 are shown for clarity) are the measurement results. The black and yellow lines are the theoretical $F_{EDL}$ curves with and without considering systematic error parameters, respectively. The $c$ in the legend are the best fitted results, and the corresponding $\kappa^{-1}$ are 30.1, 22.3, 17.9, 15.8, 14.3, and 9.65 nm, respectively. The measured $k'$ is around 0.24 N m$^{-1}$. Fitted results for (**b2**) $\psi$ and (**b3**) $\sigma$ without considering systematic error parameters ($\psi'$ and $\sigma'$). (**b4**) SEM images of the used SS on the cantilever. The measured $R'$ is 11.0 μm. Roughness determined using AFM imaging for (**b5**) SS and (**b6**) SW. (**b7**) The sectional plots corresponding to the lines in **b5**–**b6**. (**b8**) Relative height distributions with best Gaussian fits measured in **b5**–**b6**.

(**c1**) Correction results of interaction forces between a $Si_3N_4$ AFM nanoscale tip and a SW in $Li^+$ (LiCl) solutions at pH 10 measured by AFM nanoscale tip technique. The data points (only 1/2 are shown for clarity) are the measurement results. The black lines are the theoretical $F_{EDL}$ curves with considering systematic error parameters. The $c$ in the legend are the best fitted results, and the corresponding $\kappa^{-1}$ are 21.2, 14.5, 9.71, 6.87, 4.34, and 3.07 nm, respectively. The measured $k'$ is around 0.20 N m$^{-1}$. Fitted results for (**c2**) $\psi$ and (**c3**) $\sigma$ at different given $R$ ($\theta_0$ = 39.8°) without considering systematic error parameters $\Delta D$ (= 0) and $E_k$ (= 1). (**c4**) A representative SEM image of an AFM nanoscale tip. (**c5**) The schematic of the tip geometry.

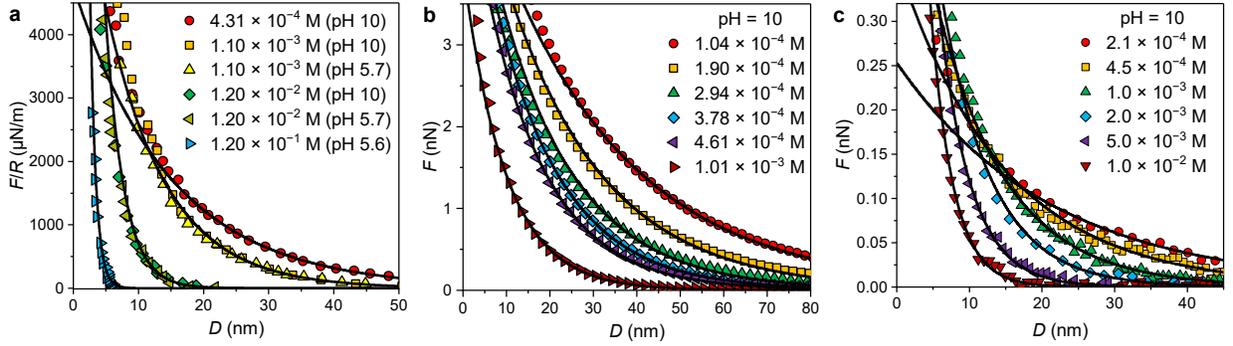

Fig. 18 Correction results in normal $xy$-axis for the three surface force measurement techniques: (**a**) SFA, (**b**) AFM microsphere probe technique, and (**c**) AFM nanoscale tip technique.

### 6.3.2  Correcting systematic errors for AFM microsphere probe technique

For AFM microsphere probe technique, interaction forces between silica microspheres (SS) and silica wafers (SW) are measured. As explained in section 2.6, the $\psi/\sigma$ of silica increases sharply with increasing pH. Therefore, to ensure that the $\psi/\sigma$ is large enough, the experiments are carried out in 10$^{-4}$–10$^{-3}$ M Li$^+$ (LiCl) solutions at pH 10. The experimental data (data points) and correction results (black lines) in logarithmic $y$-axis are shown in Fig. 17b1 (in normal $xy$-axis in Fig. 18b).

Fig. 17b1 shows that the theoretical $F_{EDL}$ curves after considering systematic error parameters at all $c$ are in good agreement with the experimental force curves, indicating the correctness of the correction method and results. The measured $R'$ is 11.0 μm (Fig. 17b4) and the measured $k'$ is 0.24 N m$^{-1}$. While the obtained systematic error parameters are $\Delta D$ = 21.0 nm and $E_k/E_R$ = 0.65.

As shown in Fig. 17b2–b3, the $|\psi'|$ and $|\sigma'|$ (misinterpreted $\psi$ and $\sigma$, fitted results for $\psi$ and $\sigma$ without considering systematic error parameters) for silica at all $c$ (10$^{-4}$–10$^{-3}$ M) at pH 10 are < 100 mV (30–70 mV) and 2–3 mC m$^{-2}$, respectively, which are much smaller than the real situation but similar to results in the literature (see sections 2.6, 1.2, and 5.1).

Fig. 17b5–b6 show the typical roughness of the SS and SW, and the $\sigma_{RMS}$ for SS and SW are 6.6 and 0.4 nm, respectively. Fig. 17b7 shows the sectional plots corresponding to the lines in Fig.

17b5–b6, and the $H_{p-v}$ for SS and SW are about 29.5 and 1.8 nm, respectively. Fig. 17b8 shows the relative height distributions with best Gaussian fits for SS and SW measured in Fig. 17b5–b6. The parameter $\sigma_{std}$ represents the standard deviation, and $4\sigma_{std}$ for SS and SW are 22.1 and 1.6 nm, respectively. The sums of $H_{p-v}$ (31.3 nm) and $4\sigma_{std}$ (23.7 nm) for SS and SW are close to $\Delta D$ but much larger than that of $\sigma_{RMS}$ (7.0 nm), indicating that it is $H_{p-v}$ that really determines $\Delta D$, not $\sigma_{RMS}$, as highlighted in section 4.1. The obtained $\Delta D$ is as large as 21.0 nm, indicating that the effects of nanoscale surface roughness are significant.

The whole $E_k/E_R = 0.65$, indicating that $E_k$ and $E_R$ together cause the experimental force curves to be 65% of the theoretical ones (without considering systematic error parameters). This can be explained by the measurement errors of $R$ and $k$ within the reported ranges of systematic errors in section 4.1.

### 6.3.3  Correcting systematic errors for AFM nanoscale tip technique

For AFM nanoscale tip technique, interaction forces between AFM nanoscale tips and SW are measured. The tip material is silicon nitride ($Si_3N_4$). As explained in section 2.6, the $\psi/\sigma$ of $Si_3N_4$ and silica increase sharply with increasing pH. Similar to section 6.3.2, to ensure that the $\psi/\sigma$ is large enough, the experiments are carried out in $10^{-4}$–$10^{-2}$ M $Li^+$ (LiCl) solutions at pH 10. The experimental data (data points) and correction results (black lines) in logarithmic $y$-axis are shown in Fig. 17c1 (in normal $xy$-axis in Fig. 18c).

Fig. 17c1 shows that the theoretical $F_{EDL}$ curves after considering systematic error parameters at all $c$ are in good agreement with the experimental force curves, indicating the correctness of the correction method and results. The nominal $R'$ is 20 nm and the measured $k'$ is 0.20 N m$^{-1}$. According to SI section 3, $\theta_0 = 39.8°$ is obtained (set $E_\theta$ (= $\theta'/\theta_0$) ≡ 1). The rest obtained systematic error parameters are $\Delta D = 4.7$ nm, $E_R = 0.61$ ($R_0 = 33$ nm), and $E_k = 1.05$.

As shown in Fig. 17c2–c3, the $\psi'$ and $\sigma'$ without considering systematic error parameters $\Delta D$ (= 0) and $E_k$ (= 1) at different given $R$ (the nominal $R'$ (20 nm), the corrected $R_0$ (33 nm), and an over-estimated $R''$ (45 nm); set $\theta_0 = 39.8°$) were calculated. The $\psi'$ and $\sigma'$ change greatly with different given $R$, even exceed the allowable range of calculation when using the nominal value 20 nm.

Despite the nanoscale of the tip, the obtained $\Delta D$ is still as large as 4.7 nm. The $H_{p-v}$ and $4\sigma_{std}$ for SW (Fig. 17b7–b8) are 1.8 and 1.6 nm, respectively, which are smaller than $\Delta D$. This indicates that the surface roughness originates from both the SW and the tip end.

## 6.4 Related notes about correction of systematic errors

For experimental systems with different surface characteristics, the sources and characteristics of systematic errors may be complicated. The identification of systematic errors in section 4.1 and the theoretical and experimental methods constructed in sections 6.1 and 6.2 are applicable to the patterned and representative experimental systems.

It therefore should be noted that the constructed theoretical and experimental methods have some restrictions. For materials with excessive roughness and very irregular shapes, like natural mineral particles (montmorillonite [88], illite [89], talc [90], calcite [91,92], etc.), direct surface force measurements and theoretical fitting calculations are not recommended (only served as a semi-quantitative or qualitative characterization) because of the excessively large deviation from the ideal situation (i.e., smooth surfaces and regular geometries).

The main prerequisite for the correction methods is to ensure that $\psi/\sigma$ is sufficiently large. For many materials with variable charge, this can be realized by setting appropriate pHs (section 2.6). While for conductors, this may also be achieved by applying a high external voltage [35,93].

An important advantage of the correction methods is that, in fact, we do not need to determine the actual surface geometry parameters ($R$ and $\theta$), surface roughness (e.g., $\sigma_{RMS}$, $H_{p-v}$, or $\Delta D$) and spring constant ($k$). The actual interpretation results ($\psi/\sigma/H$) can be directly obtained after the correction procedures. Besides, the accuracy and reliability of the interpretation results will be dramatically improved due to the correction of systematic errors.

At appropriate $c$ and pHs, the magnitude of $F_{vdW}$ can dominate over $F_{EDL}$. Based on the characteristics of $F_{vdW}$ (section 3.2), the systematic error parameters can also possibly be calculated and corrected by comparing the difference between the calculated theoretical $F_{vdW}$ curves and the experimental $F_{vdW}$ curves. However, since $F_{vdW}$ is generally very weak relative to $F_{EDL}$, the theoretical method based on $F_{vdW}$ will be not as accurate as that based on $F_{EDL}$.

## 7    Conclusion and Perspective

The main purpose of this work is to point out the pervasive anomalous results and provide corresponding reasonable explanations in measurements and interpretations of DLVO forces.

Through the elaborations in above sections, we have first identified the pervasive anomalous results mainly including: (1) the fitted $\psi/\sigma$ is normally extremely small ($\psi$ can be close to or (much) smaller than $\psi_\zeta$ (zeta potential)) and varies greatly; (2) the fitted $\psi/\sigma$ can exceed the allowable range of calculation; and (3) the measured $F_{vdW}$ and the fitted $H$ vary greatly. Then, we have reasonably explained the pervasive anomalous results in the literature and further speculated that, the pervasive anomalous results are existing but not noticed and questioned owing to the two important aspects: (1) the pervasive unreasonable understandings of EDL theory and (2) the commonly neglected systematic errors.

Consequently, we can naturally summarize that, in the last few decades since the advent of surface force measurement techniques, the relevant studies on the measurements and interpretations of DLVO forces do not seem to meet our expectations (e.g., to validate and promote the EDL theory).

The prerequisite for achieving the research goals is to establish reasonable theoretical hypotheses and experimental methods. Therefore, On the measurements and interpretations of DLVO forces, it is necessary to apply the EDL theory and consider the systematic errors carefully and appropriately, as suggested in this review.

Finally, we call for re-examination and re-analysis of related experimental results and theoretical understandings by careful consideration of the EDL theory and systematic errors. On these bases, we can interpret the experimental results properly and promote the development of EDL theory, colloid and interface science, and many related fields.

**Acknowledgments**

The first author would like to thank Prof. Markus Valtiner, Plinio Maroni, Hongbo Zeng, Michal Borkovec, Yajing Kan and his students, Manfred Heuberger, and engineers from Bruker for stimulating discussions and potential help.

**Funding:** This work was supported by the National Key R&D Program of China (2023YFD1900300).

**Author Contributions:** Bo Feng: Conceptualization; Data curation; Formal analysis; Investigation; Methodology; Validation; Visualization; Writing – original draft; Writing – review & editing. Xiantang Liu: Validation; Writing – review & editing. Xinmin Liu: Validation. Yingli Li: Validation. Hang Li: Funding acquisition; Project administration; Supervision.

**Conflict of Interest:** The authors have no conflicts to disclose.

**Supplemental Information**

# Revisiting the measurements and interpretations of DLVO forces

Bo Feng[1], Xiantang Liu[2], Xinmin Liu[1], Yingli Li[1], Hang Li[1]*

[1]*Chongqing Key Laboratory of Soil Multi-Scale Interfacial Processes, College of Resources and Environment, Southwest University, Chongqing 400715, P.R. China*
[2]*State Key Laboratory of Pollution Control and Resource Reuse, School of the Environment, Nanjing University, Nanjing 210023, P.R. China*

*Corresponding author at: Chongqing Key Laboratory of Soil Multi-Scale Interfacial Processes, College of Resources and Environment, Southwest University, Chongqing 400715, PR China.
E-mail address: lihangswu@163.com, lihang88@swu.edu.cn (H. Li).
Phone: 086-13883320589, 086-023-68251504.

Bo Feng (https://orcid.org/0000-0002-9861-3256)
Xiantang Liu (https://orcid.org/0009-0001-6120-9137)
Hang Li (https://orcid.org/0000-0002-8486-6631)

## 1 Calculations of the ratios of $\psi/\psi_\delta$ varying with $\psi_\delta$ and $\sigma/\sigma_\delta$ varying with $\sigma_\delta$

Based on Eqs. (2) and (3), we can obtain the relationship between $\psi$ and $\psi_\delta$ (set $x = \delta$):

$$\psi_\delta = \frac{4k_BT}{q}\operatorname{atanh}\left[\tanh\left(\frac{q\psi}{4k_BT}\right)e^{-\kappa\delta}\right] \Leftrightarrow \psi = \frac{4k_BT}{q}\operatorname{atanh}\left[\tanh\left(\frac{q\psi_\delta}{4k_BT}\right)e^{\kappa\delta}\right] \quad (S1)$$

and the ratios of $\psi/\psi_\delta$ can be obtained accordingly.

Based on Eq. (4), we have the relationship between $\sigma_\delta$ and $\psi_\delta$ (set $x = \delta$):

$$\sigma_\delta = \frac{2\kappa\varepsilon\varepsilon_0 k_BT}{q}\sinh\left(\frac{q\psi_\delta}{2k_BT}\right) \quad (S2)$$

Further substituting Eqs. (S1) and (S2) into Eq. (7), we can obtain the relationship between $\sigma$ and $\sigma_\delta$, and the ratios of $\sigma/\sigma_\delta$ can be obtained accordingly.

## 2 Calculations of the effects of systematic errors

As depicted in section 4.1, three systematic error parameters, $\Delta D$, $E_R$, and $E_k$, should be used to quantify the effects of systematic errors for c–c, s–f, and s–s models (Fig. 1d–f), while four systematic error parameters, $\Delta D$, $E_R$, $E_\theta$, and $E_k$, should be used for cf–f model (Fig. 1g).

(1) $F_{EDL}$ for c–c, s–f, and s–s models

Based on Eq. (6), the original theoretical $F_{EDL}$ curves without considering systematic error parameters can be expressed as

$$F_{EDL(The)} = At_1t_2\kappa R'e^{-\kappa D} \Leftrightarrow \ln F_{e(The)} = -\kappa D + \ln(At_1t_2\kappa R') \quad (S3)$$

While the final new theoretical $F_{EDL}$ curves with considering systematic error parameters $\Delta D$, $E_R$, and $E_k$, which are consistent with the experimental $F_{EDL}$ ($F_{EDL(Exp)}$) curves, should be expressed as

$$F_{EDL(The)}' = E_kAt_1t_2\kappa R'/E_R\, e^{-\kappa(D+\Delta D)} \Leftrightarrow \ln F_{EDL(The)}' = -\kappa(D+\Delta D) + \ln(E_kAt_1t_2\kappa R'/E_R) \quad (S4)$$

If we neglect the systematic errors, i.e., mistakenly believe that the theoretical $F_{EDL}$ curves without considering systematic error parameters (Eq. (S3)) and the experimental $F_{EDL}$ curves (represented by Eq. (S4)) are equal, the systematic errors will influence the interpretation of $t$ (determined by $\psi/\sigma$). Assuming $t_1 = t_2$ and by setting the actual $t$ (denoted by $t_0$, and $t_0$ is determined by the actual $\psi/\sigma$, $\psi_0/\sigma_0$) and the misinterpreted $t$ (denoted by $t'$, and $t'$ is determined by the misinterpreted $\psi/\sigma$, $\psi'/\sigma'$), we have

$$At'^2\kappa R'e^{-\kappa D} = E_kAt_0^2\kappa R'/E_R\, e^{-\kappa(D+\Delta D)} \quad (S5)$$

The relationship between $t'$ and $t_0$ can immediately be obtained by

$$t' = t_0 \sqrt{\frac{E_k}{E_R e^{\kappa \Delta D}}} \tag{S6}$$

Substituting Eqs. (3) and (7) into Eq. (S6), the relationship between $\psi'/\sigma'$ and $\psi_0/\sigma_0$ can be obtained, and the ratios of $\psi'/\psi_0$ or $\sigma'/\sigma_0$ varying with $\Delta D$, $E_R - 1$, or $E_k - 1$ at different $\psi_0/\sigma_0$ and $c$ can be obtained accordingly.

(2) $F_{EDL}$ for cf–f model

Based on Eq. (9), the original theoretical $F_{EDL}$ curves without considering systematic error parameters can be expressed as

$$\begin{aligned} F_{EDL(The)} &= A t_1 t_2 \left( \frac{1}{2} \kappa^2 R'^2 + \tan^2 \theta' + \kappa R' \tan \theta' \right) e^{-\kappa D} \\ \Leftrightarrow \ln F_{EDL(The)} &= -\kappa D + \ln \left[ A t_1 t_2 \left( \frac{1}{2} \kappa^2 R'^2 + \tan^2 \theta' + \kappa R' \tan \theta' \right) \right] \end{aligned} \tag{S7}$$

While the final new theoretical $F_{EDL}$ curves with considering systematic error parameters $\Delta D$, $E_R$, $E_\theta$, and $E_k$, which are consistent with the experimental $F_{EDL}$ ($F_{EDL(Exp)}$) curves, should be expressed as

$$\begin{aligned} F_{EDL(The)}' &= E_k A t_1 t_2 \left[ \frac{1}{2} \kappa^2 \left( R'/E_R \right)^2 + \tan^2 \left( \theta'/E_\theta \right) + \kappa \left( R'/E_R \right) \tan \left( \theta'/E_\theta \right) \right] e^{-\kappa (D + \Delta D)} \\ \Leftrightarrow \ln F_{EDL(The)}' &= -\kappa (D + \Delta D) + \ln \left\{ E_k A t_1 t_2 \left[ \frac{1}{2} \kappa^2 \left( R'/E_R \right)^2 + \tan^2 \left( \theta'/E_\theta \right) + \kappa \left( R'/E_R \right) \tan \left( \theta'/E_\theta \right) \right] \right\} \end{aligned} \tag{S8}$$

If we neglect the systematic errors, i.e., mistakenly believe that the theoretical $F_{EDL}$ curves without considering systematic error parameters (Eq. (S7)) and the experimental $F_{EDL}$ curves (represented by Eq. (S8)) are equal, the systematic errors will influence the interpretation of $t$ (determined by $\psi/\sigma$). Assuming $t_1 = t_2$ and by setting the actual $t$ (denoted by $t_0$, and $t_0$ is determined by the actual $\psi/\sigma$, $\psi_0/\sigma_0$) and the misinterpreted $t$ (denoted by $t'$, and $t'$ is determined by the misinterpreted $\psi/\sigma$, $\psi'/\sigma'$), we have

$$\begin{aligned} A t'^2 \left( \frac{1}{2} \kappa^2 R'^2 + \tan^2 \theta' + \kappa R' \tan \theta' \right) e^{-\kappa D} = \\ E_k A t_0^2 \left[ \frac{1}{2} \kappa^2 \left( R'/E_R \right)^2 + \tan^2 \left( \theta'/E_\theta \right) + \kappa \left( R'/E_R \right) \tan \left( \theta'/E_\theta \right) \right] e^{-\kappa (D + \Delta D)} \end{aligned} \tag{S9}$$

The relationship between $t'$ and $t_0$ can immediately be obtained by

$$t' = t_0 \sqrt{\frac{E_k\left[\frac{1}{2}\kappa^2\left(R'/E_R\right)^2 + \tan^2\left(\theta'/E_\theta\right) + \kappa\left(R'/E_R\right)\tan\left(\theta'/E_\theta\right)\right]}{e^{\kappa\Delta D}\left(\frac{1}{2}\kappa^2 R'^2 + \tan^2\theta' + \kappa R'\tan\theta'\right)}} \quad (S10)$$

Substituting Eqs. (3) and (7) into Eq. (S10), the relationship between $\psi'/\sigma'$ and $\psi_0/\sigma_0$ can be obtained, and the ratios of $\psi'/\psi_0$ or $\sigma'/\sigma_0$ varying with $\Delta D$, $E_R - 1$, $E_\theta - 1$, or $E_k - 1$ at different $\psi_0/\sigma_0$ and $c$ can be obtained accordingly.

(3) $F_{vdW}$ for c–c, s–f, and s–s models

Based on Eq. (8), the original theoretical $F_{vdW}$ curves without considering systematic error parameters can be expressed as

$$F_{vdW(The)} = -\frac{HR'}{6D^2} \Leftrightarrow \log\left[-F_{vdW(The)}\right] = -2\log D + \log\frac{HR'}{6} \quad (S11)$$

While the final new theoretical $F_{vdW}$ curves with considering systematic error parameters $\Delta D$, $E_R$, and $E_k$, which are consistent with the experimental $F_{vdW}$ ($F_{vdW(Exp)}$) curves, should be expressed as

$$F_{vdW(The)}' = -\frac{E_k HR'}{6(D+\Delta D)^2 E_R} \Leftrightarrow \log\left[-F_{vdW(The)}'\right] = -2\log(D+\Delta D) + \log\frac{HR'}{6} + \log\frac{E_k}{E_R} \quad (S12)$$

If we neglect the systematic errors, i.e., mistakenly believe that the theoretical $F_{vdW}$ curves without considering systematic error parameters (Eq. (S11)) and the experimental $F_{vdW}$ curve (represented by Eq. (S12)) are equal, the systematic errors will influence the interpretation of $H$. By setting the actual $H$, $H_0$, and the misinterpreted $H$, $H'$, we have

$$-\frac{H'R'}{6D^2} = -\frac{E_k H_0 R'}{6(D+\Delta D)^2 E_R} \quad (S13)$$

The ratios of $H'/H_0$ varying with $D$ at different $\Delta D$, $E_R - 1$, or $E_k - 1$ can immediately be obtained by

$$\frac{H'}{H_0} = \frac{D^2 E_k}{(D+\Delta D)^2 E_R} \quad (S14)$$

(4) $F_{vdW}$ for cf–f model

Based on Eq. (10), the original theoretical $F_{vdW}$ curves without considering systematic error parameters can be expressed as

$$F_{vdW(The)} = -\frac{H}{6}\left(\frac{\tan^2\theta'}{D} + \frac{R'\tan\theta'}{D^2} + \frac{R'^2}{D^3}\right) \tag{S15}$$

While the final new theoretical $F_{vdW}$ curves with considering systematic error parameters $\Delta D$, $E_R$, $E_\theta$, and $E_k$, which are consistent with the experimental $F_{vdW}$ ($F_{vdW(Exp)}$) curves, should be expressed as

$$F_{vdW(The)}' = -\frac{E_k H}{6}\left[\frac{\tan^2(\theta'/E_\theta)}{D+\Delta D} + \frac{(R'/E_R)\tan(\theta'/E_\theta)}{(D+\Delta D)^2} + \frac{(R'/E_R)^2}{(D+\Delta D)^3}\right] \tag{S16}$$

If we neglect the systematic errors, i.e., mistakenly believe that the theoretical $F_{vdW}$ curves without considering systematic error parameters (Eq. (S15)) and the experimental $F_{vdW}$ curve (represented by Eq. (S16)) are equal, the systematic errors will influence the interpretation of $H$. By setting the actual $H$, $H_0$, and the misinterpreted $H$, $H'$, we have

$$-\frac{H'}{6}\left(\frac{\tan^2\theta'}{D} + \frac{R'\tan\theta'}{D^2} + \frac{R'^2}{D^3}\right) = \\ -\frac{E_k H_0}{6}\left[\frac{\tan^2(\theta'/E_\theta)}{D+\Delta D} + \frac{(R'/E_R)\tan(\theta'/E_\theta)}{(D+\Delta D)^2} + \frac{(R'/E_R)^2}{(D+\Delta D)^3}\right] \tag{S17}$$

The ratios of $H'/H_0$ varying with $D$ at different $\Delta D$, $E_R - 1$, $E_\theta - 1$, or $E_k - 1$ can immediately be obtained by

$$\frac{H'}{H_0} = \frac{E_k\left[\dfrac{\tan^2(\theta'/E_\theta)}{D+\Delta D} + \dfrac{(R'/E_R)\tan(\theta'/E_\theta)}{(D+\Delta D)^2} + \dfrac{(R'/E_R)^2}{(D+\Delta D)^3}\right]}{\left(\dfrac{\tan^2\theta'}{D} + \dfrac{R'\tan\theta'}{D^2} + \dfrac{R'^2}{D^3}\right)} \tag{S18}$$

## 3 Detailed experimental procedures

Based on section 6.2, the detailed experimental procedures for correcting systematic errors in the three surface force measurement techniques are depicted as follows.

(1) SFA

In section 6.3.1, the experimental data of interaction forces between mica in Li$^+$ (LiNO$_3$) solutions based on SFA are directly selected from ref. [15]. The $\psi/\sigma$ of mica can be guaranteed to be large enough at these conditions ($c$ and pH), as explained in section 2.6. The details experimental procedures can also be found in ref. [15] and elsewhere.

(2) AFM microsphere (colloidal) probe technique

In section 6.3.2, the interaction forces between silica microspheres (SS) and silica wafers (SW) are measured by AFM microsphere (colloidal) probe technique. As explained in section 2.6, the $\psi/\sigma$ of silica increases sharply with increasing pH. Therefore, to ensure that the $\psi/\sigma$ is large enough, the force measurements are carried out in $10^{-4}$–$10^{-3}$ M Li$^+$ (LiCl) solutions at pH 10.

To guarantee the reproducibility of the experimental data, the following detailed procedures are also recommended:

1) Sample preparation. AFM microsphere (colloidal) probe is made by attaching a SS to a cantilever of an AFM probe with a two-component epoxy [94]. Prior to the force measurements, the SS (on the probe) and SW should be cleaned in air-plasma for (e.g., 5 min and 10.5W), immersed in an ample 1 M pH 10 LiCl solution for > 20 min to reduce the effects of organic contaminants and other possible ions (very important!), and then rinsed with ample ultrapure water (with a resistivity of 18.2 MΩ cm). Before mounting a prepared probe in the fluid probe holder, wash the glass window of the holder with 1–5% SDS (sodium dodecyl sulfate) and then ultrapure water to avoid possible contaminants.

2) Solution preparation. Two stock solutions of 1 M LiCl and $10^{-2}$ M LiOH need to be prepared with analytical pure LiCl and LiOH. The two stock solutions and ultrapure water should be filtered with 0.22 μm syringe filters to avoid insoluble particles. The pH 10 solutions with a series of $c$ (10 ml of each is enough) can be diluted from the two stock solutions before use.

3) Engage the probe (SS) on the sample (SW). To prevent air bubbles, use a pipet to apply a small drop of fluid on the probe so that it is fully immersed before engaging. Then put the drop of fluid onto the prepared sample (SW), align the laser on the cantilever roughly, lower the probe into the drop to form a meniscus around the fluid probe holder, readjust the laser alignment and photodetector, and engage to enter the Ramp Mode in Contact Mode in Fluid (Do not enter the Scan Mode, or the probe could be damaged.).

4) Initial Ramp parameters recommended: ramp size: ~1 μm; forward velocity: 500 nm s$^{-1}$ (low forward velocity to minimize hydrodynamic effects); reverse velocity: 2 μm s$^{-1}$ (slightly high reverse velocity to save time. In fact, the hydrodynamic effects are almost undetected at this velocity.); number of samples: ≥ 1024 (at least around 1 data point per 1 nm)); Z closed loop: on; trigger mode: relative; Trig threshold: around 2 V (see below).

5) Calibrate the deflection sensitivity (DS). The Z scanner must be well calibrated first. The deflection sensitivity (DS, unit: nm/V) allows conversion from the raw photodiode signal (in V (volt)) to deflection of the cantilever (in nm), and is equal to the inverse of the slope of the force curve while the cantilever is in contact with a hard sample surface. Use the trace (forward) line as the active force curve. The "contact" region reflects the "zero distance" ($D' = 0$) of the force curve with a precision of ~0.5 nm and should be achieved by applying sufficient loading force (Trig threshold) to overcome the EDL force. The DS should be carefully calibrated and monitored, and should be recalibrated one a change is detected. Do not touch any laser position controls because the DS will change with the laser position on the cantilever.

6) Determine cantilever spring constant $k$. The apparent $k$, $k'$, of the cantilever was measured by the thermal tune method (based on measuring thermal noise) [95]. Besides, appropriate $k$ should be selected. The measured force may exceed the allowable range with a too small $k$, while the force sensitivity (nm/nN) may become very low with a too large $k$.

7) Force measurements. Use the solutions sequentially from low to high $c$. Carefully rinse the SW and SS (pick up the scanner with the holder and probe) with the solution drops three times before using each solution to avoid interference. Conduct force measurements with each solution after an equilibration time of 20 min (Don't take too long as the solution will continue to absorb carbon dioxide.). Select at least three random sites for each solution and measure 10 approach–retraction cycles per site. To replace the solution, it is highly recommended to loosen the dovetail and pick up the scanner head directly at the 1 mm withdraw distance (other operating methods may severely influence the DS).

8) Criteria for the reproducibility of the measured force curves.
   a) The DS is unchanged and the "contact" region is achieved for all force curves.
   b) Unusual long-range pull-off adhesion force usually indicates the presence of organic contaminants. The two curves of each approach–retraction cycle should overlap completely, except at high $c$ and short range where $F_{vdW}$ overcomes $F_{EDL}$ to create adhesion in retraction curve. The force curves were considered valid only if no difference was observed between every force curve at all sites.

c) Based on section 3.2, $F_{EDL}$ is linearly related to $D$ and the slope is $-\kappa$ in natural logarithmic y-axis. Therefore, the fitted slope $-\kappa$ of the $D > \kappa^{-1}$ part of the measured force curves should match the set value of $c$.

On these bases, any force curve can be used due to the excellent reproducibility.

In section 6.3.2, The SS with nominal diameter of 10 μm (Shanghai Aladdin Biochemical Technology Co., Ltd, China) are attached to the C cantilevers of NP-O tip-less probes (Bruker) [94]. The SW with 300-nm thick thermal oxide layers (Zhejiang Lijing Silicon Material Co., Ltd, China) are used. The SS and SW are cleaned in air-plasma (PDC-36G, Hefei Kejing Material Technology Co., Ltd, China). All force measurements are carried out at room temperature of 23 ± 2 °C by an AFM in Contact Mode (Dimension Icon Nanoscope V., Bruker, Camarillo, CA, USA). After the force measurements, the apparent radius $R'$ of each used SS on the cantilever (sputter coated with gold) (Fig. 17b4) are measured by a scanning electron microscope (SEM) (Phenom Pro, Thermo Fisher Scientific Inc., USA). Surface topography of SS and SW (Fig. 17b4–b5) are measured by AFM imaging in ScanAsyst mode in air by the same AFM with the SNL probes (Bruker) with a nominal tip radius ~2 nm.

(3) AFM nanoscale tip technique

In section 6.3.3, the interaction forces between AFM nanoscale tips and SW are measured by AFM nanoscale tip technique. All experimental procedures are recommended the same as above.

The tip material is silicon nitride ($Si_3N_4$). As explained in section 2.6, the $\psi/\sigma$ of $Si_3N_4$ and silica increases sharply with increasing pH. To ensure that the $\psi/\sigma$ is large enough, the experiments are carried out in $10^{-4}$–$10^{-2}$ M $Li^+$ (LiCl) solutions at pH 10.

The E cantilevers of MLCT-BIO probes (Bruker) were used. The geometry of the tip is a square pyramid and the angle between its each face and the central axis (z-axis) is 35 ± 2° (Fig. 17c4–c5) as provided from the manufacturer. The probe is mounted at a tilt angle of 12° with respect to the sample surface. Therefore, according to the calculation method in our previous work [33], $\theta_0 = 39.8 ± 2.2°$ is obtained (set $E_\theta (= \theta'/\theta_0) \equiv 1$). The tip end is smoothed into a flat surface on a clean SW in Contact Mode under ultrapure water with a large loading force prior to the experiments. This procedure is employed to match the cf–f (Fig. 1g) model which can improve measurement accuracy by producing well-defined flat surface, increasing interaction area, and reducing the effects of tip wear [33,85–87]. A representative image of the tip (sputter coated with gold) (Fig. 17c4) was taken by a SEM as above.